\documentclass{mn2e}
\usepackage{natbib}
\usepackage{graphicx}
\usepackage{threeparttable}

\begin{document}


\title{Double-double radio galaxies: further insights into the formation of the radio structures}

\author[C. Brocksopp et al.]{C. Brocksopp$^1$, C.R. Kaiser$^2$, A.P. Schoenmakers$^3$ and A.G. de Bruyn$^{3,4}$\\
$^1$ Mullard Space Science Laboratory, University College London, Holmbury St. Mary, Dorking, Surrey, RH5 6NT, UK\\
$^2$ School of Physics \& Astronomy, University of Southampton, Southampton SO17 1BJ, UK\\
$^3$ Stichting ASTRON, PO Box 2, 7990 AA Dwingeloo, The Netherlands\\
$^4$ Kapteyn Institute, PO Box 800, 9700 AV Groningen, The Netherlands\\
}

\maketitle

\begin{abstract}
Double-double radio galaxies (DDRGs) offer a unique opportunity for us to study multiple episodes of jet activity in large-scale radio sources. We use radio data from the Very Large Array and the literature to model two DDRGs, B1450+333 and B1834+620, in terms of their dynamical evolution. We find that the standard Fanaroff-Riley II model is able to explain the properties of the two outer lobes of each source, whereby the lobes are formed by ram-pressure balance of a shock at the end of the jet with the surrounding medium. The inner pairs of lobes, however, are not well-described by the standard model. Instead we interpret the inner lobes as arising from the emission of relativistic electrons within the outer lobes, which are compressed and re-accelerated by the bow-shock in front of the restarted jets and within the outer lobes. The predicted rapid progression of the inner lobes through the outer lobes requires the eventual development of a hotspot at the edge of the outer lobe, causing the DDRG ultimately to resemble a standard Fanaroff-Riley II radio galaxy. This may suggest that DDRGs are a brief, yet normal, phase of the evolution of large-scale radio galaxies.
\end{abstract}


\section{Introduction}
Fanaroff-Riley type II radio galaxies (FRII; \citealt{fr74}) are well-known for the ejection of highly collimated jets from a central nucleus. These jets subsequently inflate a pair of radio synchrotron-emitting lobes where the jets are decelerated by the ram-pressure of the intergalactic medium (IGM). ``Double-double radio galaxies'' (DDRGs) form a sub-class of FRII objects in which there are multiple pairs of lobes, all strongly coaligned to within a few degrees of each other. There are currently $\sim$ 12--17 such systems known, identified and studied by \citet{sbrlk00}, \citet{sbrl00}, \citet*{ksr00}, \citet{ksj06}, \citet*{skk06}, \citet*{ms09}, \citet*{mjk10} and \citet*{sj09}). Recently the first ``double-double radio quasar'', 4C~02.27 has been discovered (\citealt*{jsk09}), showing that the properties of DDRG are not purely phenomena of giant radio sources and may even be a common, but short-lived, phase of active galaxy evolution.

The second, inner pair of lobes in each DDRG/Q was interpreted as evidence for a second episode of jet activity, within the remnant cocoon of the original jet, as opposed to knots in an underlying jet. Support for this conclusion was increased significantly by the discovery by \citet{bks07} of a {\em third} pair of closely aligned lobes in the first known ``triple-double radio galaxy'', B0925+420. Such objects therefore give us rare and invaluable opportunities to study the duty-cycles of large-scale radio galaxies, despite their episodes of jet activity taking place on timescales of millions of years.

\citet{bks07} modelled the three pairs of lobes of B0925+420 in terms of their dynamical evolution in order to determine the permitted ranges of jet power as a function of age of the source and density of the surrounding medium, thereby testing the hypothesis that the subsequent pairs of lobes were indeed forming within the cocoon of the original jet. They found that the formation of the inner pair of lobes was indeed consistent with the outer pair having been displaced buoyantly by the ambient medium. The middle lobes were more problematic, requiring higher densities than those found within the outer lobes. An alternative model (\citealt{cb91}) was suggested, interpreting the inner and middle lobes as jet-driven bow-shocks within the outer lobes.

\begin{table*}
\caption{Angular resolutions, position angles (PA) and 1$\sigma$ noise levels in each of the new VLA images.}
\begin{tabular}{lcccccc}
\hline
\hline
Frequency   &A-array Res. (PA)  &A-array 1$\sigma$ &Time on    &B-array Res. (PA)    &B-array 1$\sigma$ & Time on\\
(GHz)       & (\arcsec) ($^{\circ}$)   &($\mu\mbox{Jy\,beam}^{-1})$ &Source (mins.)& (\arcsec) ($^{\circ}$)    &($\mu\mbox{Jy\,beam}^{-1}$)&Source (mins.)\\
\hline
{\bf B1450+333}    &                   &                        &                &                   &              &\\
1.40     & $1.7\times 1.5$\,($-12.4$)   & $217.0$&236  & $6.6\times 4.9$\,(74.8)           & $214.0$&55\\
4.86     & $0.6\times 0.5$\,($-21.9$)   & $28.9$& 117   & $1.6\times 1.5$\,($-85.9$)        & $38.5$&88\\
8.46     & $0.3\times 0.2$\,($-7.3$)    & $29.1$& 117   & $0.9\times 0.8$\,($-72.8$)        & $29.9$&89\\
\hline
{\bf B1834+620}    &                   &                        &               &                   &              &\\
1.40     & $1.9\times 1.5$\,($-8.33$)   & $212.0$& 249   & $5.9\times 4.9$\,(17.09)          & $556.0$&62\\
4.86     & $0.5\times 0.3$\,($13.89$)   & $44.1$& 116   & $2.0\times 1.5$\,($-60.8$)        & $64.6$&59\\
8.46     & $0.3\times 0.2$\,($-19.4$)   & $27.6$&98    & $1.2\times 0.9$\,($-60.8$)        & $41.1$&57\\
\hline
\end{tabular}
\label{resolution}
\end{table*}

In order to further fine-tune the mechanisms by which the observed characteristics of the lobes can be reproduced, it is necessary to model more lobes from other DDRG systems. In this paper we use Very Large Array radio data for two additional systems, B1450+333 and B1834+620, which have been the subjects of detailed studies previously (\citealt{ksj06}; \citealt{sbrl00}). We summarise previous results from the literature in the remainder of Section 1, we present the VLA observations in Section 2, the resulting total intensity, spectral index and polarisation images in Sections 3 and 4 for B1450+333 and B1834+620 respectively and then use the results as inputs for the model in Sections 5--7. We note that the modelling sections assume a WMAP cosmology, with $H_0=71\mbox{km\,s}^{-1}\mbox{Mpc}^{-1}$, $\Omega_{\mbox{m}}=0.27$, $\Omega_{\Lambda}=0.73$ (\citealt{s03}).

\subsection{B1450+333}
B1450+333 (=J1453+3308) was identified as a DDRG by \citet{sbrlk00} using data from Faint Images of the Radio Sky at 20-cm (FIRST: \citealt{bwh95}), the NRAO VLA Sky Survey (NVSS: \citealt{ccg98}) and the Westerbork Northern Sky Survey (WENSS: \citealt{rtb97}). It consists of four radio lobes distributed about a central core and only the southern, outer lobe shows tentative evidence for a weak hotspot. The southern lobes are weaker and more extended than the northern lobes and the outer lobes display an extended morphology perpendicular to, as well as along, the jet axis. \citet{ksj06} studied this object further, using a wide range of observing frequencies and resolutions. They modeled the spectra of the lobes and determined spectral ages for the northern and southern outer lobes of $\sim 47$ and 58 Myr respectively, implying a mean separation velocity of $0.036c$; the synchrotron age of the inner lobes was $\sim 2$ Myr, implying a velocity of $\sim 0.1c$. B1450+333 lies at a redshift of 0.249 (\citealt{sbrlk00}).

\subsection{B1834+620}
B1834+620 was similarly identified as a DDRG by \citet{sbrlk00, sbrl00} using data from the WENSS and NVSS. It too consists of four radio lobes about a central core, with hotspots present at the leading edges of the northern outer and southern inner lobes. The authors used the presence of these hotspots to argue in favour of the inner lobes having rapid advance velocities in a low-density environment; such a conclusion was supported by the discrepancy between the large optical emission line luminosity of the host galaxy and the relatively low radio power of the inner lobes, as well as by the model of \cite{ksr00}. B1834+620 lies at a redshift of 0.519 (\citealt{sbrl00}).

\section{Observations}


\begin{figure}
\includegraphics[width=8cm,angle=0]{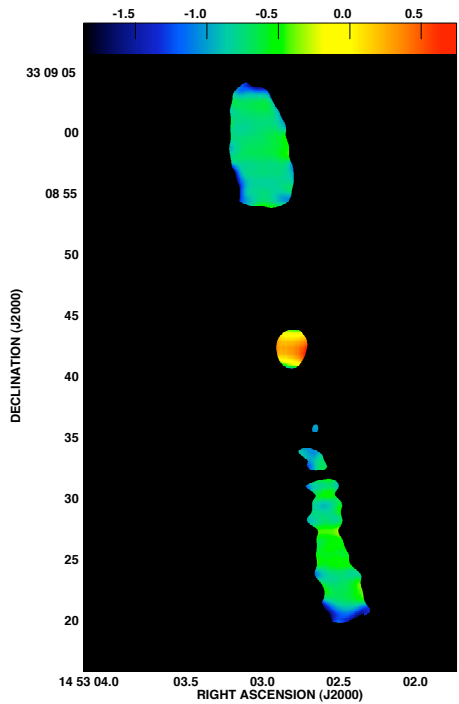}
\caption{Spectral index map created from the VLA data for the inner lobes of B1450+333. We used the B-array images at 4.86-GHz so as to ensure that no emission was resolved out and the A-array image at 1.40-GHz so as to match the resolutions as closely as possible. The higher-resolution image was reconvolved with the beam of the lower-resolution image to give an angular resolution $=1.7\times 1.5\arcsec$. The core appears to have a relatively flat spectrum and the lobes a steeper spectrum with $\alpha$ predominantly in the range $-0.5$ to $-1.5$}
\label{alpha:B1450}
\end{figure}

\begin{figure}
\includegraphics[width=8cm,angle=0]{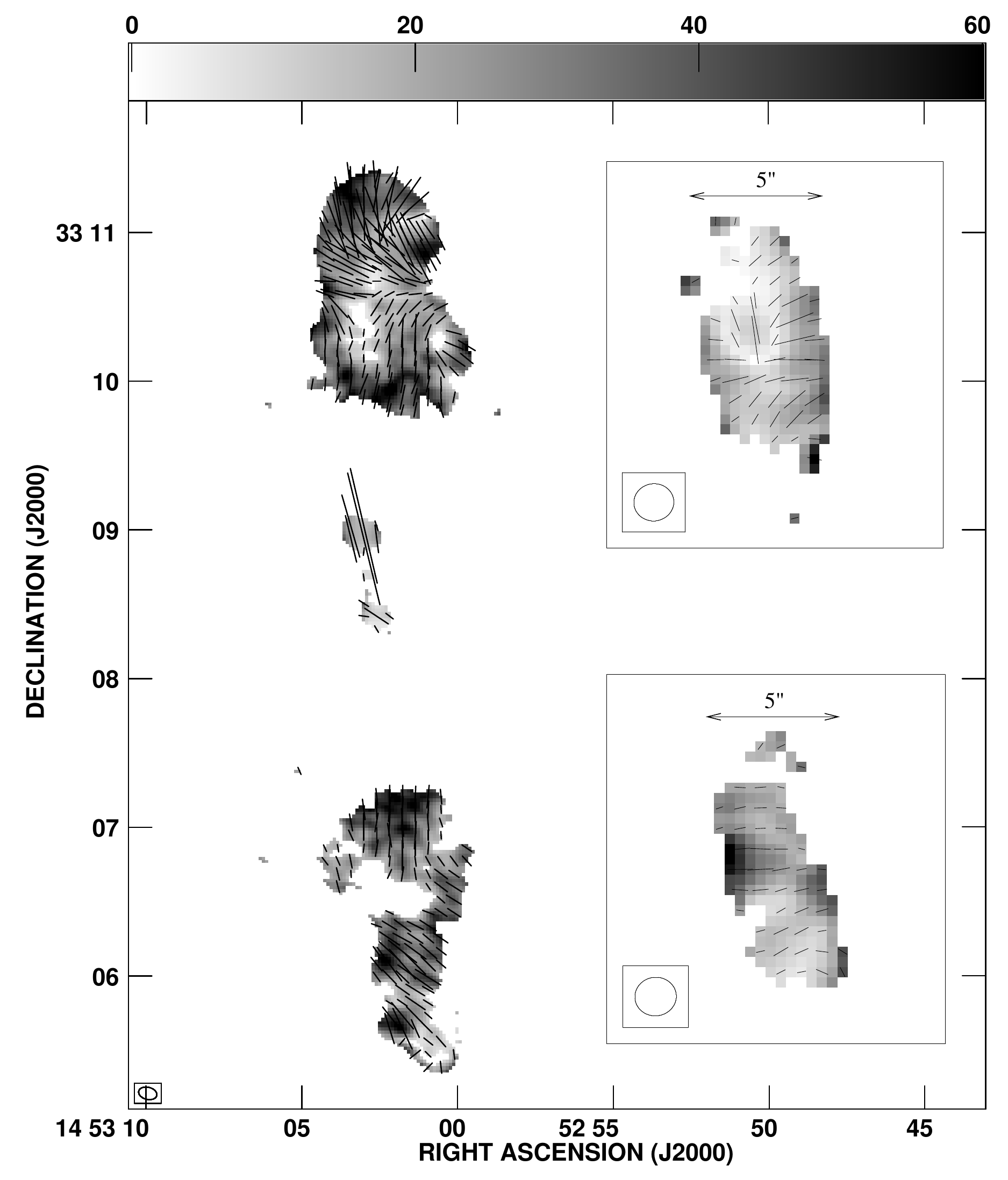}
\caption{Fractional linear polarisation (FP) images of B1450+333 using the VLA:B-array data only. FP is represented by the greyscale, which ranges from 3$\sigma$ to the map peak in each case. The ``wedge'' indicates the greyscale in the main image only. The main image is at 1.40-GHz and the inset plots of the inner lobes are at 4.86-GHz. The vectors indicate the direction of the electric field, following correction for RM (=14 rad m$^{-2}$), and their length indicates the linear polarisation ($1\arcsec = 4.81\times10^{-5}\mbox{Jy\,beam}^{-1}$). The outer lobes are very strongly polarised -- up to $\sim 60 \%$ in places, particularly around the edge of the northern lobe -- while the rest of the lobes are still typically polarised up to $\sim 20\%$. The vectors show a complicated structure to the electric field, particularly in the outer lobes and we discuss this further in the text.}
\label{FP:B1450}
\end{figure}

Radio observations of B1450+333 and B1834+620 were obtained using the VLA on 2000 October 20--21, when the array was in the highest resolution A-configuration, and 2001 April 2, when the array was in the B-configuration. Data were obtained at three frequencies -- 1.40-GHz (L-band), 4.86-GHz (C-band) and 8.46-GHz (X-band) -- in each array, as for our previous observing campaign for B0925+420 (\cite{bks07}); IFs at two adjacent frequencies were used in each continuum observation. In order to reduce potential bandwidth-smearing in the most extended image, the 1.40-GHz B-array observation of B1450+333 was obtained in the PA/PB spectral line observing mode. This gave us 8 channels of 3.125 MHz each, resulting in a 25-MHz bandwidth instead of the more usual, continuum bandwidth totalling 50 MHz. Full polarisation information was also recorded. 3C\,286 (JVAS J1331+3030) was used as the flux and polarisation calibrator while JVAS J1416+3444 and JVAS J1849+670 were used for phase calibration of B1450+333 and B1834+620 respectively and to ensure parallactic angle coverage.

The 2000/2001 VLA data were reduced using standard flagging, calibration, cleaning and imaging routines within {\sc aips} (version 31DEC04). The flux densities of 3C\,286 were obtained using the formulae of \cite{bgp77} but with the revised coefficients of Rick Perley, as is the default option in the {\sc setjy} routine. Additional analysis routines within {\sc aips} were used to measure lobe parameters and create spectral index maps. We list the angular resolutions and sensitivities of the images at all three frequencies and in both array-configurations in Table~\ref{resolution}.

We note that these VLA observations of B1450+333 have been presented by \citet{ksj06} and so we do not repeat their work here. The B1834+620 observations are previously unpublished, although a detailed study of this source has also been presented by \citet{sbrl00}.

\section{Results -- B1450+333}
B1450+333 was detected at each of the six frequency/resolution combinations. The B-array images clearly resolve the core and inner lobes, although the outer lobes are only detected in the B-array images at 1.40-GHz and much of the extended emission is resolved out in the A-array maps. The higher-frequency images hint at an extended morphology to the inner lobes, approximately aligned along the axis of the large-scale source. The morphology of this object has been studied in further detail by \citet{ksj06} and images at various frequencies can be seen there.

We used the VLA images to measure the flux, length and width of each lobe, typically using the lower-resolution images, and to create spectral index images (Fig.~\ref{alpha:B1450}); we define the spectral index, $\alpha$, in terms of the $S_{\nu}\propto\nu^{\alpha}$, where $S_{\nu}$ is the flux density at frequency $\nu$. The spectral index images use B-array data at the two higher frequencies, so as to minimise the amount of emission resolved out, and A-array data at 1.40-GHz so as to match the resolutions as closely as possible. The lobes have $\alpha$ predominently in the range $-0.5$ to $-1.5$. whereas the core is flatter and/or inverted. There appear to be some localised regions of steeper spectrum, particularly at higher frequencies. These values of $\alpha$ are consistent with those calculated for e.g. B0925+420 (\citealt{bks07}) and the lobes of more typical FRII radio galaxies, confirming that all the flux has been recovered. We use these measurements as input parameters for the modeling in Section 7.

\begin{figure*}
\hspace*{-1cm}\includegraphics[width=20cm,angle=0]{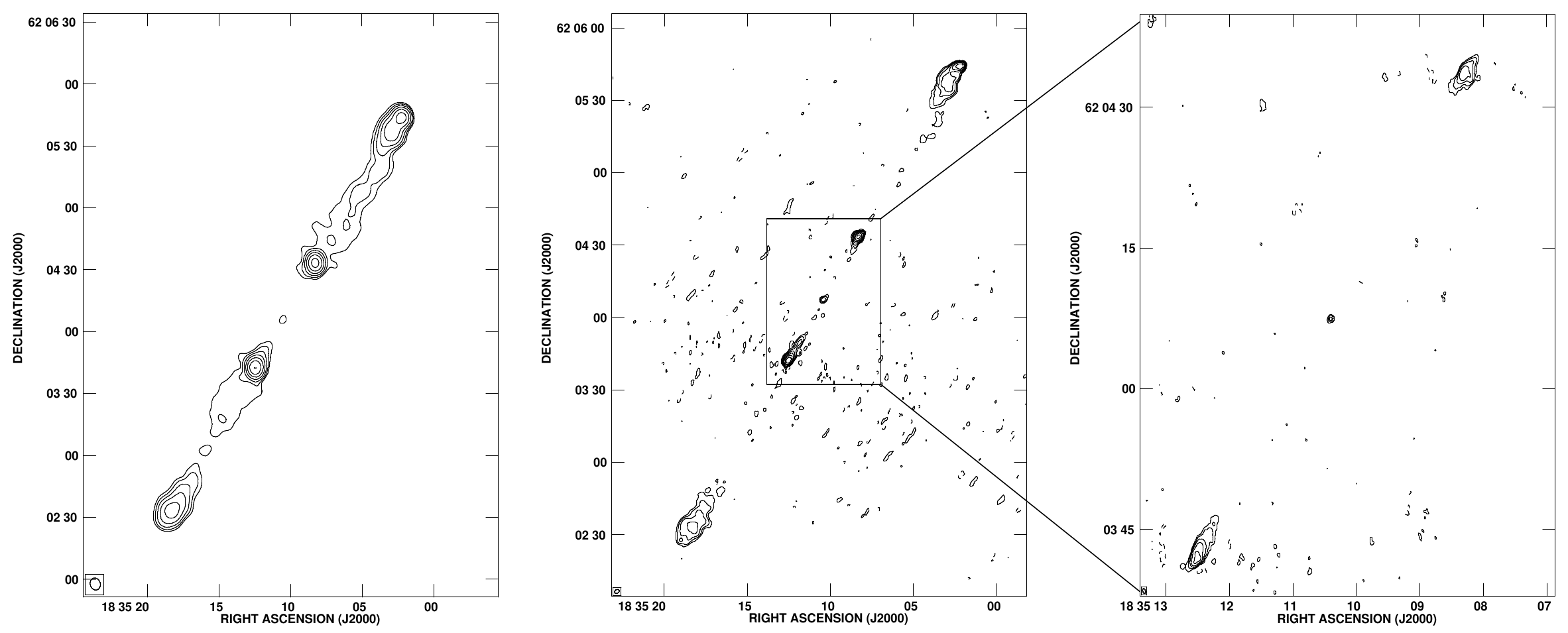}
\caption{VLA images of B1834+620 showing the two pairs of lobes. All contours are at -3, 3, 6, 12, 24, 48, 96, 192$\times \sigma$. LHS: the 1.40-GHz B-array data showing extended emission along the jet axis ($\sigma=556\,\mu\mbox{Jy\,beam}^{-1}$). Centre: 4.86 GHz B-array data showing discrete lobes without extension ($\sigma=64.6\,\mu\mbox{Jy\,beam}^{-1}$). Right: 4.86-GHz A-array image showing the inner lobes only ($\sigma=44.1\,\mu\mbox{Jy\,beam}^{-1}$).}
\label{images:B1834}
\end{figure*}


\begin{figure}
\includegraphics[width=9cm,angle=0]{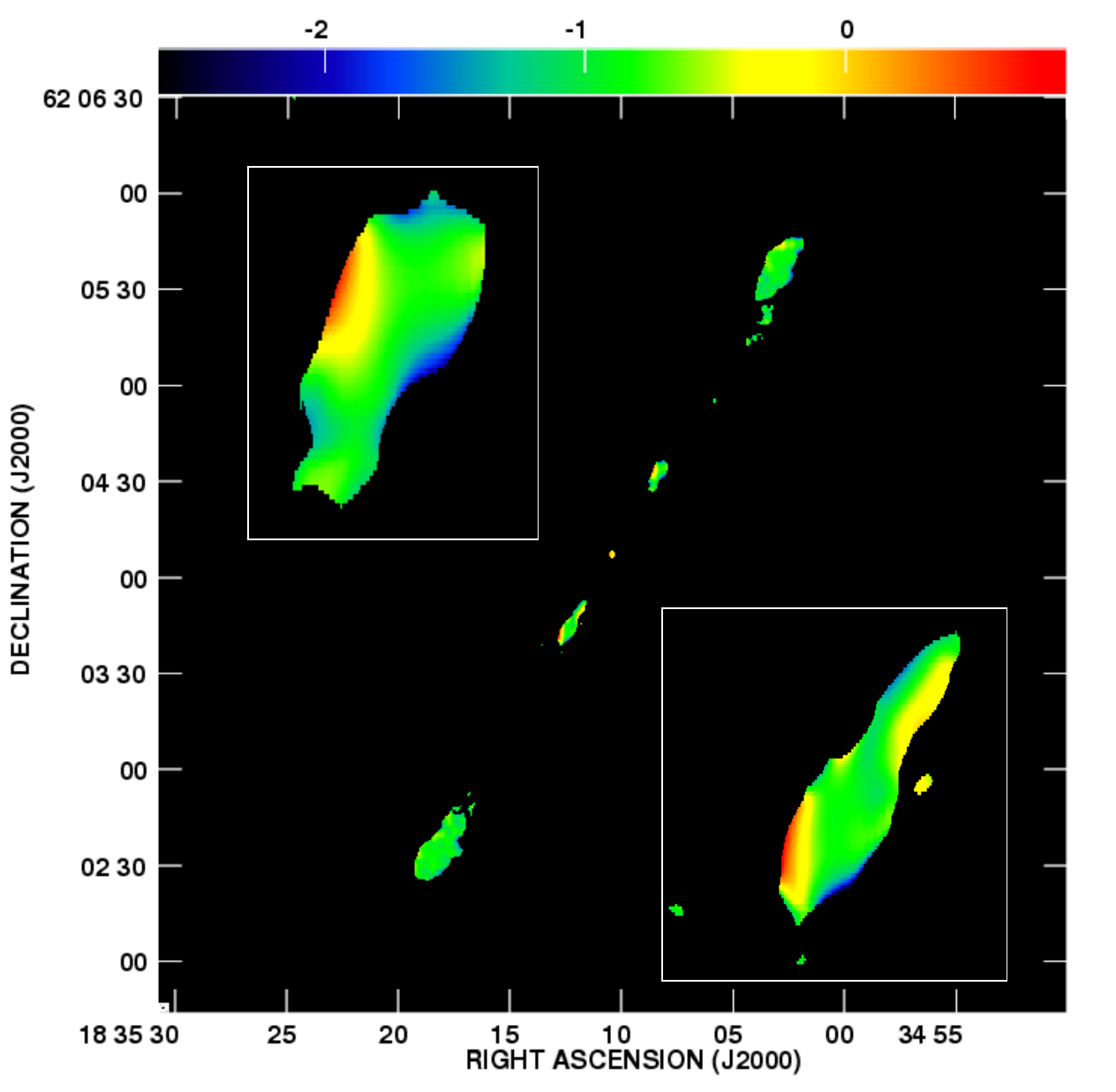}
\caption{Spectral index maps created for the inner and outer lobes of B1834+620; we used the A-array image for 1.4-GHz and the B-array image for 4.86-GHz so as to ensure that no emission was resolved out at the higher frequencies. The higher-resolution image was reconvolved with the beam of the lower-resolution image, resulting in an angular resolution $=1.9\times 1.5\arcsec$. The core appears to have a relatively flat spectrum and the lobes a steeper spectrum with $\alpha$ in the range $-0.5$ to $-2$. Inset plots show the inner lobes at higher magnification.}
\label{alpha:B1834}
\end{figure}

Using B-array polarisation images at all three frequencies, we produced a rotation measure (RM) image. Using the {\sc curval} routine within {\sc aips}, we measured typical pixel values of $\le 10$ rad m$^{-2}$, consistent with that of \citealt{ksj06}, although occasional higher pixel values resulted in a mean (using the {\sc imstat} routine within {\sc aips}) of 14 rad m$^{-2}$. Using this higher value, we derived corrections to the position angle of polarisation of 32 and 3$^{\circ}$ for 1.40- and 4.86-GHz respectively. Fig.~\ref{FP:B1450} shows the resultant fractional polarisation images with electric vectors corrected for RM; the main image is at 1.40-GHz so as to show the outer lobe structure (and at higher resolution than the polarisation images presented by \citealt{ksj06}) and the inset plots of the inner lobes are at 4.86-GHz. The electric field vectors are roughly perpendicular to the jet direction for most parts of the inner lobes, as is typical for lobes in standard radio galaxies. However there are also significant disruptions to an ordered field, particularly in the centre of the northern inner lobe. The magnetic field becomes perpendicular to the source axis in the outer regions of (in particular) the northern outer lobe. This may be caused by compression due to lobe expansion, as also suggested for the middle lobes of B0925+420 (\citealt{bks07}), which may seem surprising for a lobe which is no longer supplied with additional energy (see Section 7). The level of fractional polarisation depends on location within the lobe. The northern outer lobe showed higher levels (40--60 \%) of FP around the outer parts and lower levels (7--20\%) towards the centre. The southern outer lobe showed similar levels but with a more ``patchy'' structure, perhaps caused by a clumpy composition, consistent with the 20\% depolarisation detected by \citet{ksj06}. The inner lobes were slightly less polarised but still with the higher levels tending towards the sides of the lobes.

\section{Results -- B1834+620}
Similarly, B1834+620 was detected at each of the six frequency/resolution combinations although much of the extended emission was resolved out in the high-frequency A-array images. Images at each frequency are shown in Fig.~\ref{images:B1834}, with inset plots overlaying the right-hand panel to show the inner lobes at high resolution. All the lobes are extended along the axis of the large-scale source and the 1.40-GHz images hint at some possible emisson extending from the outer lobes back towards the inner lobes. This source has previously been studied by \cite{sbrl00} and here we extend their analysis using new, higher-resolution images.

We used the VLA images to obtain fluxes, widths and lengths of the lobes, again typically using the lower-resolution images, and to create spectral index images (Fig.~\ref{alpha:B1834}). The values of $\alpha$ appear reasonable and of the order $\sim -1$ for lobes and flatter in the core; however, the spectrum in the 4.86-GHz/8.46-GHz image appears unexpectedly steep ($-3$ to $-4$) in places suggesting that some of the 8.46-GHz emission has been resolved out. We therefore use the fluxes of \cite{sbrl00} in Section 7.

\begin{figure}
\includegraphics[width=8cm,angle=0]{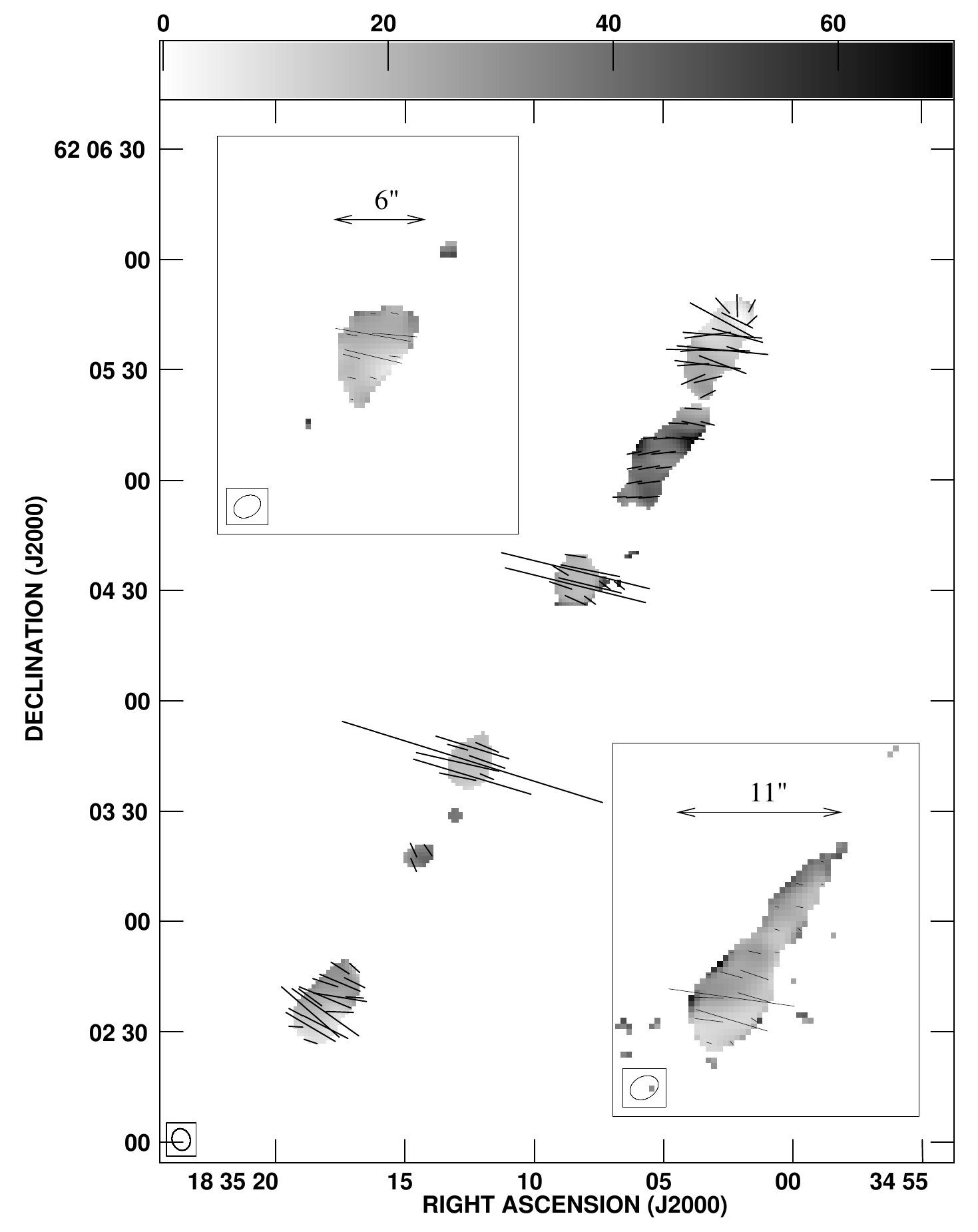}
\caption{Fractional linear polarisation (FP) images of B1834+620 using the VLA:B-array data only; FP is represented by the greyscale, which ranges from 3$\sigma$ to the map peak in each case. The ``wedge'' indicates the greyscale in the main image only. The main image is at 1.40-GHz and the inset plots of the inner lobes are at 4.86-GHz. The vectors indicate the direction of the electric field, following correction for RM (=55 rad m$^{-2}$), and their length indicates the linear polarisation ($1\arcsec = 4.81\times10^{-5}\mbox{Jy\,beam}^{-1}$). The outer lobes are strongly polarised, typically up to $\sim 20\%$. The electric vectors are perpendicular to the source axis in all lobes, with the exception of the hotspot in the north outer lobe.}
\label{FP:B1834}
\end{figure}

The discrepancy affecting the spectral index image also appears to affect our estimates for the rotation measure; we obtain a very low value of 0.4 rad m$^{-2}$ which is inconsistent with the higher value of 55--60 obtained by \cite{sbrl00}. Given the much greater number of frequencies used by these authors, and the lower resolution of their images, we adopt their value for correction of our images. We note that unpublished data from the Westerbork Synthesis Radio Telescope further support this higher value of rotation measure (de Bruyn et al. in prep.). The resultant fractional polarisation image at 1.40-GHz is plotted in Fig.~\ref{FP:B1834} with inset plots showing the inner lobes at 4.86-GHz. Contrary to B1450+333, the electric vectors are approximately perpendicular to the jet axis in each of the four lobes, with the main region of disruption taking place around the leading edge of the northern outer lobe. This location corresponds with the bright hotspot, an uncommon feature of DDRG objects, although common to detect in more conventional FRII galaxies. The degree of fractional polarisation is less extreme in this object, typically around 20\% in all lobes, although possibly a little higher in parts of the northern outer lobe which, again, is likely to be related to the hotspot.

\section{A model for the outer lobes}
\label{KDA}

The standard model for the evolution of the large-scale structure inflated by jets in FRII-type radio galaxies was first proposed by \citet{ps74} and \citet{br74}. The momentum transported along the jet is balanced by the thermal pressure and ram pressure of the receding medium in the vicinity of the host galaxy. The momentum balance leads to the formation of a strong shock at the end of the supersonic jet. After passing through the jet shock the jet material inflates a lobe or cocoon of low density, but high pressure, gas. This material protects the jet flow from turbulent disruption through contact with the higher density ambient medium. 

\citet{sf91} showed that an initially overpressured jet will come into pressure equilibrium with its own lobe. For a jet propagating into an ambient medium with a power-law density distribution, i.e.
\begin{equation}
\rho_{\rm x} = \rho \left( \frac{r}{a} \right)^{-\beta},
\label{dens}
\end{equation}
where $r$ is the distance from the centre of the distribution and $\rho$ and $a$ are constants, the growth of the lobe will be self-similar (\citealt{ka96b}). The size of the lobe is then related to the age of the jet flow, $t$, by a power-law. 

We can also calculate the radio luminosity of the lobe of a given size (e.g. \citealt{kda97a,br00,mk02}). This takes into account the cumulative effects of energy losses of the radiating electrons due to the adiabatic expansion of the lobe, the emission of synchrotron radiation and the inverse Compton scattering of Cosmic Microwave Background (CMB) photons. Here we use the model of the source dynamics by \citet{ka96b} and that for the radio synchrotron emission of the lobe by \citet{kda97a}. This combined model was recently summarised and discussed in \citet{kb07} (see also \citealt{kb08}) and we will use the model in the form presented there.

In the radio maps of the two DDRGs there is little or no evidence for active hotspots in the outer lobes. This suggests that they are no longer supplied with energy by active jets. Here we assume that the energy supply to the outer lobes only stopped recently in terms of the total source age, $t$. The dynamics of the lobes are then still governed by the same equations as during the active jet phase lasting a time $t_{\rm j} < t$. However, no freshly relativistic accelerated electrons enter the lobes after $t_{\rm j} + t_{\rm t}$, where $t_{\rm t}$ is the time it takes the last jet material ejected by the central active galactic nucleus (AGN) at $t_{\rm j}$ to reach the end of the still growing lobe (\citealt{ksr00,bks07}). We note that the parameter $t_{\rm j}$ did not appear in the standard model, referenced above, but is introduced here as a free parameter.

The combined model involves a number of parameters. These can be divided into three groups. Set parameters are fixed to plausible values by considering the underlying physics or observations independent of the radio observations. The value of observed parameters are derived from the observational radio data of individual DDRGs discussed in this paper. Free parameters are those derived by making the model self-consistent and taking into account all observational information. Table \ref{para} summarises all parameters and we discuss the more important ones below.

\begin{table*}
\begin{tabular}{llcc}
\hline
 & Adiabatic index of lobe material & $\Gamma_{\rm l}$ & 5/3\\
 & Adiabatic index of external medium & $\Gamma_{\rm x}$ & 5/3\\
 & Adiabatic index of magnetic field & $\Gamma_{\rm B}$ & 4/3\\
 & Minimum of initial electron Lorentz factor &  $\gamma_{\rm min}$ & 1\\
 Set & Maximum of initial electron Lorentz factor & $\gamma_{\rm max}$ & $10^8$\\
 & Viewing angle & $\theta$ & $90^{\circ}$\\
 & Exponent of external density distribution & $\beta$ & 0, 1.5\\
 & Core radius of external density distribution & $a$ & 2\,kpc\\
 & Energy density of the CMB at $z=0$ & $u_{\rm CMB}$ & $4\times 10^{-14}$\,J\,m$^{-3}$\\
 & Ratio of energy densities of non-radiating particles and relativistic electrons & $k$ & 0, 100\\[1.5ex]
 & Length of lobe & $D$ & \\
 & Aspect ratio of lobe & $A$ & \\
 Observed & Observing frequency & $\nu$ & \\
 & Radio luminosity density & $L_{\nu}$ & \\
 & Cosmological redshift & $z$ & \\
 & Exponent of initial power-law energy distribution of relativistic electrons & $m$ & \\[1.5ex]
 & Jet power & $Q$\\
 Free & External density at core radius & $\rho$ & \\
 & Source age & $t$ & \\
 & Duration of jet flow & $t_{\rm j}$ & \\
 \hline
 \end{tabular}
 \caption{Summary of model parameters for the standard FRII model. The notation is that used in \citet{kb07}.}
 \label{para}
 \end{table*}
 
 The adiabatic index of the lobe material depends on its composition. For lobes filled only with magnetic fields and a relativistic pair plasma, we would expect a relativistic equation of state with $\Gamma_{\rm l} = 4/3$. However, in the following we find that the energy density of the lobes of B\,1450+333, and to a much lesser extent those of B\,1834+620, may contain a significant contribution from other, non-radiating particles, i.e. $k\ne0$ (where $k$ is the ratio of the energy densities of non-radiating particles and relativistic electrons). Also, if the power-law energy distribution of the electrons does indeed extend down to Lorentz factors of around unity, then most particles will not be highly relativistic and hence we set $\Gamma _{\rm l} =5/3$. The limits of the initial electron energy distribution do not strongly influence the results as long as $m \sim 2$, but for steeper power-laws the lower energy cut-off may become more important. This is the case for the inner lobes of the DDRGs discussed here and we will return to this point in Section \ref{stand}.
 
 The way in which the ratio of the energy density of the magnetic field in the lobe and that of all particles, $R$, is defined in \citet{kda97a} and \citet{kb07} implies that the relativistic electrons and the magnetic field are not fulfilling the minimum energy condition at any point in time when $k\ne0$. However the magnetised plasma in the lobes of FRII-type sources appears to be close to minimum energy conditions \citep[e.g.][]{chh05}. Hence we enforce minimum energy conditions by setting $R=3 / \left[ 4 \left( k+1 \right) \right]$. In this way, all model equations involving $R$ and $k$ remain valid.

The outer lobes of the DDRGs discussed in this paper are very large. The model therefore predicts the sources to be old. Any inclination of the lobes towards our line of sight, implying a viewing angle $\theta < 90^{\circ}$, would result in projection with the actual lobe sizes and their ages increasing even further. However, this is a small effect as long as $\theta > 45^{\circ}$. While the observational data do not rule out projection effects, we assume that the lobes are lying in the plane of the sky. 

The density distribution of the ambient gas is described by equation (\ref{dens}). The model therefore only depends on the combination $\rho a^{\beta}$, but not on $a$ and $\rho$ individually. Here we choose to fix $a$ and adopt $\rho$ as a free parameter. Any change suggested by the model for $\rho$ may be interpreted as a change of $a$ instead as long as the combination $\rho a^{\beta}$ retains the value suggested by the model. The exponent of the power-law $\beta$ is set to 1.5, appropriate for a wide range of gaseous environments from elliptical galaxies to galaxy clusters \citep[e.g.][]{fmo05}. We also investigate the possibility that the outer radio lobes extend beyond the local environment of the host galaxy or cluster into a more uniform Inter Galactic Medium (IGM). In this case we set $\beta =0$. 

The length of the lobes and the ratio of their length and their width, the lobe aspect ratio $A$, can be determined from radio observations with sufficient resolution to resolve the lobe perpendicular to the jet axis. However, at high observing frequencies, allowing sufficient spatial resolution, the lobes are often not detected all the way from the hotspots to the core. While we can still determine the length of the lobes, it is difficult to measure their width. Most measurements of $A$ must therefore be viewed as upper limits. 

Measuring the total radio luminosity density at a given observing frequency requires low spatial resolution to avoid resolving out flux on the largest scales. At the same time we require sufficient resolution to distinguish emission from the inner and outer lobes. For both sources we find measured fluxes at a range of frequencies that avoid both problems. 

The power-law exponent for the initial energy distribution of the relativistic electrons may, in principle, be determined from the observed spectral slope of the lobes at low frequencies where radiative energy losses of the electrons have not yet significantly changed the energy distribution. This is straightforward in the case of young sources like the inner lobes and below we will use the value for $m$ derived from their observed spectral slope. A similar determination of $m$ is difficult to achieve for older sources like the outer lobes of the DDRGs \citep[e.g.][]{ksj06}. We show below that the observed spectra of the outer lobes are consistent with a comparatively flat energy distribution with $m=2$. 

We use the observational data to constrain the values of four free parameters, the jet power, $Q$, the normalisation of the density distribution of the ambient medium, $\rho$, the source age, $t$, and the duration of the jet flow, $t_{\rm j}$. Assuming that we have $n$ measurements of the radio flux at the same number of frequencies spread out over a large range plus the length of the lobe, then the degrees of freedom of the model are given by $n-3$. For the inner lobes we assume the jets are still supplying these structures with energy and so the number of free parameters is reduced to three, because $t=t_{\rm j}$. Also, we show below that the spectra of the inner lobes are consistent with power-laws. Hence the individual flux measurements are not independent and they can only be counted as a single measurement. For the inner lobes we therefore have only two observational constraints, but three free parameters, resulting in relations between pairs of free parameters instead of specific values for each of them.

\section{Models for the inner lobes}

The spectra of the inner lobes, up to an observed frequency of at least 8.46 GHz, are consistent with a power-law description. This implies that the electrons emitting in the observed frequency range have not suffered significant radiative energy losses since they were accelerated to relativistic velocities. The inner lobes must therefore be young. This requirement is consistent with our assumption that the inner lobes represent a second activity phase of the jet flow from the central AGN after the energy supply to the outer lobes shuts down. The inner lobes then cannot be older than a few $10^7$\,years, because for an older age the emission from the outer lobes should have faded below the fluxes at the higher observed frequencies. Given the observed close alignment of the inner lobes with the outer lobes, we assume that the inner lobes are fully contained within the volumes of the outer lobes. We propose two different interpretations for the existence and observed properties of the inner lobes. 

\subsection{Standard FRII model}
\label{stand}

The first interpretation is that the same model used to explain the outer lobes also applies to the inner lobes. The jet flows inflating the inner lobes must then end in strong shocks giving rise to particle acceleration. The observations do not show much evidence for luminous radio hotspots which could be identified as the jet shocks. However, given our limited spatial resolution, we cannot rule out the presence of compact structures with an enhanced radio luminosity at the end of the inner lobes. 

The model is identical to that developed in Section \ref{KDA} for the outer lobes. However, we can make some simplifications. We assume that the inner lobes are currently still supplied with energy by active jets. Hence $t_{\rm j} =t$ and we reduce the number of free parameters by one. Also, the power-law shape of the observed spectrum implies insignificant radiative energy losses of the relativistic electrons. The inner lobes will be well described by the much simplified model in the adiabatic regime as described in \citet{kb07}. For an observed lobe length of $D$ and radio luminosity density $L_{\nu}$ it is straightforward to derive relations between the source age of the inner lobes, $t_{\rm i}$, and the jet power, $Q$,
\begin{equation}
t_{\rm i} = \left[ \frac{ \left( 1+ \epsilon \right)^4 D^{3 \left( m+ 1 \right)} L_{\nu}^4}{f_{\rm L}^4 f_{\rm p}^{m+1} c_1^{\left( 5 - \beta \right) \left( m+ 1 \right) / 3}} \right]^{1/ \left( m + 5 \right)}Q^{-1},
\label{innert}
\end{equation}
where $\epsilon$, $f_{\rm L}$, $f_{\rm p}$ and $c_1$ are constants defined in \citet{kb07}. We can also find a relation between the scale density of the ambient gas, $\rho_{\rm i}$, and the jet power,
\begin{equation}
\rho_{\rm i} = \left( \frac{c_1}{D} \right)^{5-\beta} a^{-\beta} Q t_{\rm i}^3.
\label{innerQ}
\end{equation}

The observed power-law spectra have steep slopes and in the absence of significant radiative energy losses these steep slopes translate into large values for the exponent of the initial energy distribution of the relativistic electrons, $m$. Large $m$ may indicate inefficient particle acceleration and they make the model results more dependent on the low energy cut-off of the electron energy distribution as most of the energy of the electrons is stored in the low energy end of the distribution. However, our conclusions below do not depend on our assumptions for $\gamma_{\rm min}$. In general, this standard FRII interpretation of the inner lobes requires substantially higher mass densities within the outer lobes than can be explained with the material transported along the jet flows that inflated these structures. 

\subsection{Bow shock model}
\label{bowmod}

For the alternative model we assume that the density and pressure inside the outer lobes is insufficient to significantly slow down the advance of the inner, young jets. In this case, the inner jets do not develop very strong shocks at their ends or inflate substantial lobes. We do not expect the ends of the jets to be sites of efficient particle acceleration and so the inner jets do not directly produce much radio emission. However, they push aside and compress the contents of the outer lobes. The advance of the inner jets is fast and therefore leads to the formation of a bow shock within the outer lobes. This bow shock can re-energise the relativistic electrons and also strengthens the magnetic field by compressing it. We identify the radio synchrotron emission observed from the inner lobes with that produced by the shocked and compressed outer lobe material. The model was discussed briefly in \citet{bks07}, where we also point out that bow shock structures, but no lobes directly inflated by jets, are found in numerical simulations of restarting jets \citep{cb91}. Here we expand the description of the model to see whether our observations of B\,1450+333 and B\,1834+620 are consistent with its predictions. A very similar model was recently developed independently by \citet{ssb08} to describe the inner lobes of the DDRG PKS~B1545$-$321.

In the restframe of the bow shock propagating through the outer lobe ahead of the inner jet, the momentum supplied by the jet must be balanced by the momentum of the receding lobe material. We must also include the pressures of the jet material, $p_{\rm j}$, and of the lobe material, $p_{\rm l}$, in this balance and so \citep[e.g.][]{gb94}
\begin{equation}
\left( \gamma _{\rm jb}^{2} \beta_{\rm jb}^2 w_{\rm j} + p_{\rm j}\right) A_{\rm j} = \left( \gamma_{\rm b}^2 \beta_{\rm b}^2 w_{\rm l} + p_{\rm l} \right) A_{\rm b},
\end{equation}
where $\beta_{\rm j}$ is the speed of the bow shock in the restframe of the outer lobe material and $\beta_{\rm jb}$ is the speed of the jet material in the restframe of the bow shock. $\gamma_{\rm b}$ and $\gamma _{\rm jb}$ are the corresponding Lorentz factors. From the relativistic transformation of velocities we get $\gamma _{\rm jb}^{2} \beta_{\rm jb}^2 = \gamma_{\rm b}^2 \beta_{\rm j}^2 \left( \beta_{\rm j} - \beta _{\rm b} \right)^2$. The jet has a cross-section of $A_{\rm j}$ and the bow shock has a surface area of $A_{\rm b}$.

The relativistic enthalpy $w$ is defined as $w=e+p$ with $e$ the internal energy density, including the restmass energy density, and $p$ the pressure. For simplicity we assume that all gases can be described by an equation of state of the form $p=\left( \Gamma - 1 \right) \left( e - \rho c^2 \right)$, where $\Gamma$ is the ratio of specific heats and $\rho$ the restmass density. It is convenient to introduce the ratio ${\cal R} = \rho c^2 / \left(w - \rho c^2 \right)$. For large ${\cal R}$ the energy density of the gas is dominated by its restmass energy density, while for small values of ${\cal R}$ the gas is relativistic. In \citet{bks07} we only considered the latter case. We can now write
\begin{eqnarray}
& & \left[ \gamma_{\rm b}^2 \beta_{\rm j}^2 \left( \beta_{\rm j} - \beta_{\rm b} \right)^2 \frac{\Gamma_{\rm j}}{\Gamma _{\rm j} - 1} \left( {\cal R}\,_{\rm j} +1 \right) + 1 \right] p_{\rm j} A_{\rm j}  \nonumber \\
& =& \left[\gamma_{\rm b}^2 \beta_{\rm b}^2 \frac{\Gamma_{\rm l}}{\Gamma_{\rm l} -1} \left( {\cal R}\,_{\rm l} +1 \right) + 1 \right] p_{\rm l} A_{\rm b}.
\label{betab}
\end{eqnarray}

The energy flux along the jet is
\begin{equation}
Q = \left( \gamma_{\rm j}^2 w_{\rm j} - \rho_{\rm j} c^2 \right) \beta_{\rm j} c A_{\rm j}.
\end{equation}
Substituting for $w_{\rm j}$ and $\rho_{\rm j}$ in terms of $p_{\rm j}$ and ${\cal R}\,_{\rm j}$ and solving for the jet cross-section gives
\begin{equation}
A_{\rm j} = \frac{\Gamma _{\rm j} - 1}{\Gamma_{\rm j}} \frac{Q}{\beta_{\rm j} c p_{\rm j} \left[ \gamma _{\rm j}^2 + {\cal R}\,_{\rm j} \left( \gamma _{\rm j}^2 - 1 \right) \right]}.
\label{area}
\end{equation}
Note that upon substituting this expression for $A_{\rm j}$ into equation (\ref{betab}), the jet pressure terms cancel. 

We now turn to the properties of the material in the outer lobe and find an expression for ${\cal R}\,_{\rm l}$. The jet inflated this lobe during a time $t_{\rm j}$. If its volume is $V_{\rm l}$ and it only contains material that was supplied by the jet, i.e. mixing across the lobe surface can be neglected, then
\begin{equation}
\rho _{\rm l} = \frac{\dot{M} t_{\rm j}}{V_{\rm l}},
\label{outerdensity}
\end{equation}
where $\dot{M}$ is the restmass transport rate of the jet. $\dot{M}$ is given by
\begin{equation}
\dot{M} =  \rho_{\rm j} \beta_{\rm j} c A_{\rm j}.
\end{equation}
Substituting for $\rho_{\rm j}$ in terms of ${\cal R}\,_{\rm j}$ and $p_{\rm j}$ as well as substituting for $A_{\rm j}$ we find
\begin{equation}
\dot{M} = \frac{Q}{c^2} \frac{{\cal R}\,_{\rm j}}{\gamma_{\rm j}^2 +{\cal R}\,_{\rm j} \left( \gamma_{\rm j}^2 - 1 \right)}.
\label{masstransport}
\end{equation}
Since 
\begin{equation}
{\cal R}\,_{\rm l} = \frac{\Gamma _{\rm l} - 1}{\Gamma_{\rm l}} \frac{\rho_{\rm l} c^2}{p_{\rm l}},
\end{equation}
we can now find the value of ${\cal R}\,_{\rm l}$, if we know $p_{\rm l}$ and the other quantities appearing in equations (\ref{betab}) and (\ref{area}). We obtain $p_{\rm l}$, $Q$, $t_{\rm j}$ and $V_{\rm l}$ from the model for the outer lobe, while we can use the observed geometry of the inner lobes to constrain the surface area of the bow shock, $A_{\rm b}$. Assuming either a matter dominated, ${\cal R}\,_{\rm j} \gg 1$, relativistic, ${\cal R}\,_{\rm j} \ll 1$, or balanced, ${\cal R}\,_{\rm j} \sim 1$, jet we can then find the speed of the bow shock, $\beta_{\rm b}$, as a function of the Lorentz factor of the jet, $\gamma_{\rm j}$, from the implicit equation (\ref{betab}).

The material in the outer lobe passes through the bow shock driven by the inner jet. In this process, the mass, momentum and energy of the lobe material must be conserved. This leads to the continuity equations across a relativistic shock \citep[e.g.][]{ll87},
\begin{eqnarray}
\beta_{\rm b} \gamma _{\rm b} \rho_{\rm l}  & = & \beta_{\rm s} \gamma _{\rm s} \rho_{\rm s}, \nonumber\\
\beta_{\rm b}^2 \gamma_{\rm b}^2 w_{\rm l} + p_{\rm l} & = & \beta_{\rm s}^2 \gamma_{\rm s}^2 w_{\rm s} + p_{\rm s}, \\
\beta_{\rm b} \gamma_{\rm b}^2 w_{\rm l} & = & \beta_{\rm s} \gamma_{\rm s}^2 w_{\rm s}, \nonumber
\end{eqnarray}
where a subscript `s' denotes quantities describing the shocked material behind the bow shock. All velocities are measured in the restframe of the bow shock. If we assume that the ratio of specific heats of the material in the outer lobe, $\Gamma_{\rm l}$, does not change during the shock passage, then we can re-arrange the set of equations above to give
\begin{equation}
\beta_{\rm b} \gamma_{\rm b} \left[ 1 - \frac{\gamma_{\rm b}}{\gamma_{\rm s}} + \frac{\Gamma_{\rm l}}{\Gamma_{\rm l} -1} \beta_{\rm s} \left( \beta_{\rm b} -\beta_{\rm s} \right) \gamma_{\rm b} \gamma_{\rm s} \right] = \frac{\beta_{\rm b} \gamma_{\rm b} -\beta_{\rm s} \gamma_{\rm s}}{1 + {\cal R}\,_{\rm l}}
\label{betas}
\end{equation}
and
\begin{equation}
p_{\rm s} = \left[ \frac{\Gamma_{\rm l}}{\Gamma_{\rm l} -1} \beta_{\rm b} \gamma_{\rm b}^2 \left( \beta_{\rm b} - \beta_{\rm s} \right) \left( 1 + {\cal R}\,_{\rm l} \right) + 1 \right] p_{\rm l}.
\label{ps}
\end{equation}
From the implicit equation (\ref{betas}) we obtain the velocity of the shocked gas with respect to the bow shock, $\beta_{\rm s}$. Substituting this into equation (\ref{ps}) we arrive at the pressure of the shocked gas $p_{\rm s}$ which may be compared to the pressure estimated from the radio synchrotron emission of the inner lobes by using the minimum energy condition.

\section{Application of the models}

We now apply the models detailed above to the observational data on B\,1450+333 and B\,1834+620. 

\subsection{B\,1450+333}

\begin{table}
\begin{threeparttable}
\begin{tabular}{lcccc}
\hline
\hline
{\bf B1450+333:}& \multicolumn{2}{c}{North} & \multicolumn{2}{c}{South}\\
& Outer & Inner & Outer & Inner\\
Lobe length, $D$ / kpc & 635\tnote{a} & 79.8\tnote{b} & 781\tnote{a} & 83.7\tnote{b}\\
Aspect ratio, $A$ & 1.8\tnote{c} & 3.9\tnote{d} & 3\tnote{c} & 5\tnote{d}\\
\hline
& \multicolumn{4}{c}{Flux densities / mJy}\\
\hline
0.240\,GHz & 1055 & 123 & 602 & 48\\
0.334\,GHz & 902	& 96	& 563 & 47\\
0.605\,GHz & 606	& 69	& 370 & 29\\
1.287\,GHz & 272 & 37 & 164 & 17\\
1.365\,GHz & -- & 38 & -- & 14\\
1.40\,GHz & 258 & -- & 145 & --\\
4.86\,GHz & 71 & 16 & 32 & 4.9\\
8.46\,GHz & -- & 9.5 & -- &2.6\\
\hline
\end{tabular}
\caption{Observed properties of the four lobes in the DDRG B\,1450+333 used in the modelling. Additional flux densities have been taken from the compilation by \citet{ksj06}.}
\label{1450obs}
\begin{tablenotes}
\item [a] VLA in B-array at 1.40\,GHz
\item [b] VLA in A-array at 4.86\,GHz
\item [c] GMRT at 0.605\,GHz \citep{ksj06}
\item [d] VLA in A-array at 1.40\,GHz
\end{tablenotes}
\end{threeparttable}
\end{table}

The radio observations used to constrain the models for the inner and outer lobes of B\,1450+333 are summarised in Table \ref{1450obs}. We follow \citet{ksj06} and assume a typical error for the flux measurements of 7\% at all frequencies, except at 0.240\,GHz where the error is assumed to be 15\%.

\subsubsection{Outer lobes}

\begin{table*}
\begin{tabular}{lcccccc}
\hline
\hline
 & \multicolumn{2}{c}{$k=0$, $\beta=1.5$} & \multicolumn{2}{c}{$k=0$, $\beta =0$} & \multicolumn{2}{c}{$k=100$, $\beta=1.5$}\\
 & North & South &North & South & North & South\\
 \hline
 $Q / 10^{37}\,{\rm W}$ & 7.0 & 8.5 & 5.6 & 6.5 & 410 & 500\\
 $t / {\rm Myr}$ & 53 & 43 & 55 & 48 & 52 & 43\\
 $t_j / {\rm Myr}$ & 48 & 34 & 50 & 39 & 46 & 33\\
 $\rho / 10^{-24}\,{\rm kg}\,{\rm m}^{-3}$ & 3.0 & 0.58 & $1.3\times 10^{-3}$ & $2.2\times10^{-4}$ & 170 & 34\\
 $\chi ^2$ & 0.68 & 0.86 & 0.35 & 0.73 & 0.70 & 0.75\\[1ex]
 $p_{\rm l} / 10^{-15}\,{\rm J}\,{\rm m}^{-3}$  & 3.1 & 1.5 & 3.5 & 1.7 & 180 & 88\\
 $B_{\rm l} / {\rm nT}$ & 0.071 & 0.049 & 0.075 & 0.052 & 0.071 & 0.049\\
 $\rho_{\rm l} \left( \xi = 1 \right) / 10^{-24}\,{\rm kg}\,{\rm m}^{-3}$ & $6.4 \times 10^{-7}$ & $8.2 \times 10^{-7}$ & $5.4 \times 10^{-7}$ & $7.2 \times 10^{-7}$ & $3.7 \times 10^{-5}$ & $4.8 \times 10^{-5}$ \\
 \hline
 $\rho_{\rm i} / 10^{-24}\,{\rm kg}\,{\rm m}^{-3}$ & 0.023 & $2.0 \times 10^{-4}$ & 0.036 & $2.0 \times 10^{-4}$ & $7.4 \times 10^{-4}$ & $3.6 \times 10^{-6}$\\
 $t_{\rm i} / {\rm Myr}$ & 2.4 & 0.41 & 2.9 & 0.54 & 0.19 & 0.031\\
\hline 
\end{tabular}
 \caption{Model parameters for the outer lobes of B\,1450+333 for the model spectra with the best fit. The model parameters for the inner lobes for the standard FRII model, $t_{\rm i}$ and $\rho_{\rm i}$, evolving inside the respective outer lobes are also given.}
 \label{1450mod}
 \end{table*}

\begin{figure}
\includegraphics[width=8.45cm]{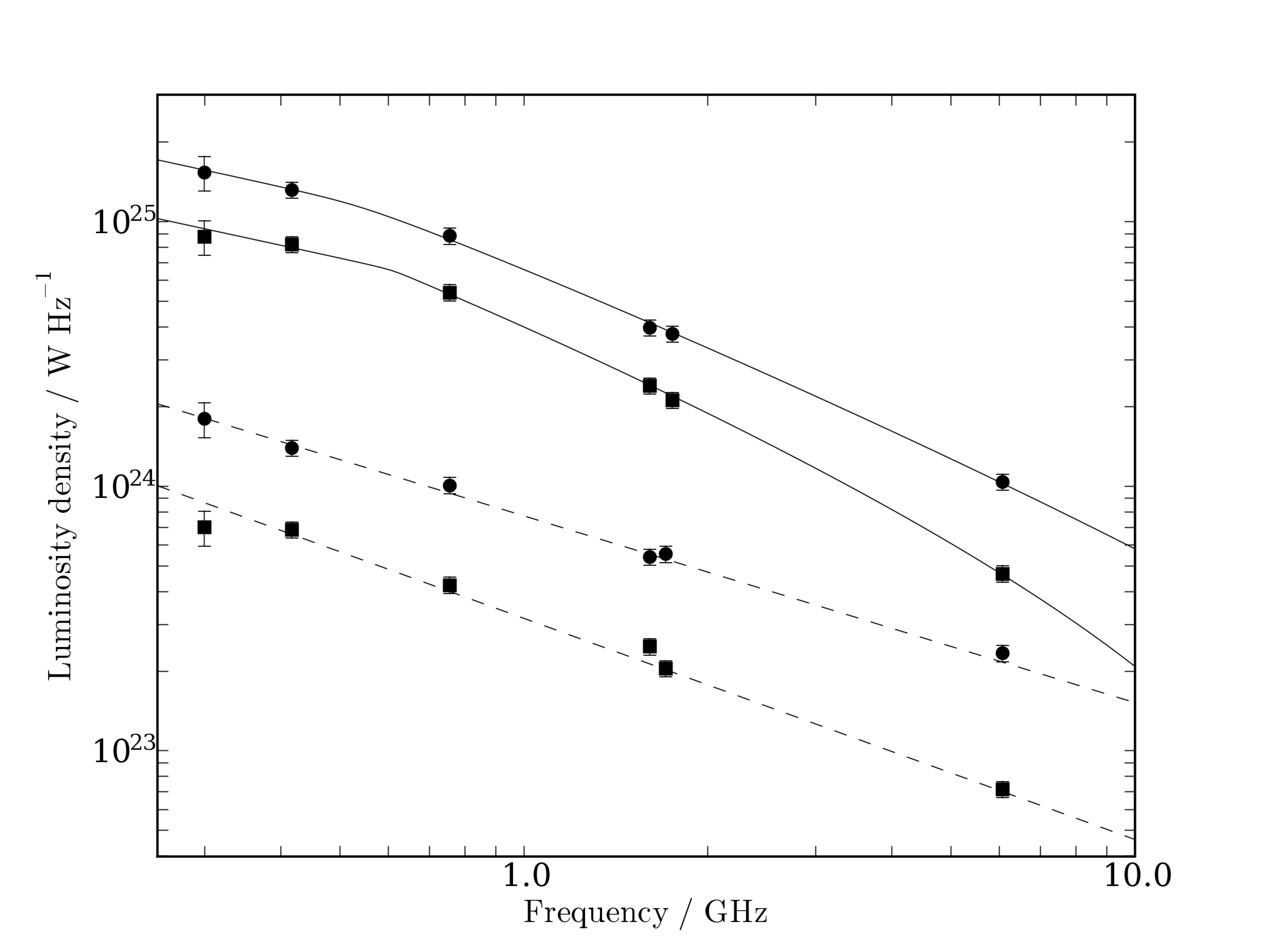}
\caption{Comparison of model spectra with observations for B\,1450+333 in the source restframe. Circles show the data for the northern lobes, squares are for the southern lobes. The upper data points and lines are associated with the outer lobes while the lower data points and lines show the inner lobes. The solid lines show the model predictions for the outer lobes, the dashed lines show the power-law fits to the spectra of the inner lobes.}
\label{1450spec}
\end{figure}

For the outer lobes we have flux measurements at six frequencies for each lobe. According to the discussion in Section \ref{KDA} this implies that we have three degrees of freedom in the model fitting. We investigate three individual models. The first assumes a power-law density distribution the jets expand into with $\beta=1.5$ and no energy in any non-radiating particles in the lobes, i.e. $k=0$. We then allow for non-radiating particles in the lobe by setting $k=100$. Finally we assume that the outer radio lobes extend beyond the confines of the atmosphere of the host galaxy into a uniform intergalactic medium with $\beta=0$. For this model we implicitly assume that the lobes have extended beyond the atmosphere of their host galaxy for a significant fraction of their age. Table \ref{1450mod} shows the fitting results for these three models and Fig.~\ref{1450spec} shows the model spectrum for the case $\beta=1.5$ and $k=100$ compared to the observations. 

In all three cases we get very good agreement of the model spectra with the observational data, despite the fact that the observed spectral slope at high frequencies is considerably steeper than the $\alpha=0.5$ expected from a model with $m=2$ in the absence of radiative energy losses of the electrons. The fact that $\chi^2 < 1$ for all three cases suggests that our assumed errors of the flux measurements, 15\% at 0.240\,GHz and 7\% at the other frequencies, are too conservative. Also, based on the $\chi^2$ values alone, none of the models is formally preferred over the others. 

While we cannot formally reject any of the models, a closer inspection of the model results shows that both models with $k=0$ require a very low gas density in the source environment. We can convert the gas density measured from X-ray observations of galaxy groups by \citet{jph07} to our choice of $\beta=1.5$ and $a=2$\,kpc. The lowest density in this sample is that of NGC\,3607 with $\rho \sim 8 \times 10^{-23}$\,kg\,m$^{-3}$ followed by NGC\,720 with $\rho \sim 1.40 \times 10^{-22}$\,kg\,m$^{-3}$. The densities required by our model with $\beta =1.5$ are at least a factor 25 lower. Also, for $\beta =0$, when we assume that the outer lobes extend beyond the gaseous atmosphere of the host galaxy, the inferred density is at least a factor 14 below the critical density of the universe at this redshift. We consider it therefore unlikely that the models with $k=0$ accurately describe the outer lobes of B\,1450+333. 

We experimented with changing the values of the various set model parameters to arrive at a more realistic density for the source environment. The only model parameter forcing a significant change of $\rho$ is $k$, the ratio of the energy stored in non-radiating particles inside the outer lobes. Setting $k=100$, we found an acceptable fit with a density for the host environment consistent with the findings of \citet{jph07} for galaxy groups. The age of the outer lobes does not change much for this model because the observed spectral shape requires significant ageing of the radiating electrons. However, the larger energy requirements force a large increase in the jet power, $Q$. While still larger values for $k$ are conceivable, given that for $k=100$ the density inferred by the model is still at the lower end of the range of the sample of \citet{jph07}, the required increase in $Q$ would put the jet power of B\,1450+333 beyond the maximum inferred for radio galaxies of type FRII \citep[e.g.][]{rs91}. Allowing $m$ to be a free parameter can help to reduce the high required values of $Q$ \citep[compare with e.g.][]{mcsk07}, although at risk of making the model too flexible. However, increasing $m$ too much results in the outer lobes becoming so young that there is insufficient time for the inner lobes to develop.

The strength of the magnetic field inferred by the model is independent of the value of $k$ while the pressure in the lobes increases substantially for larger $k$. This is just a reflection of the assumption, implicit in our model, of conditions close to equipartition inside the lobes. A given radio luminosity originating in a given lobe volume fully determines the strength of the magnetic field \citep[e.g.][]{ml94}. The magnetic field and the synchrotron emitting electrons both contribute to the lobe pressure, $p_{\rm l}$, but for $k=100$, $p_{\rm l}$ is dominated by the energy density of the non-radiating particles. 

\begin{figure}
\includegraphics[width=8.45cm]{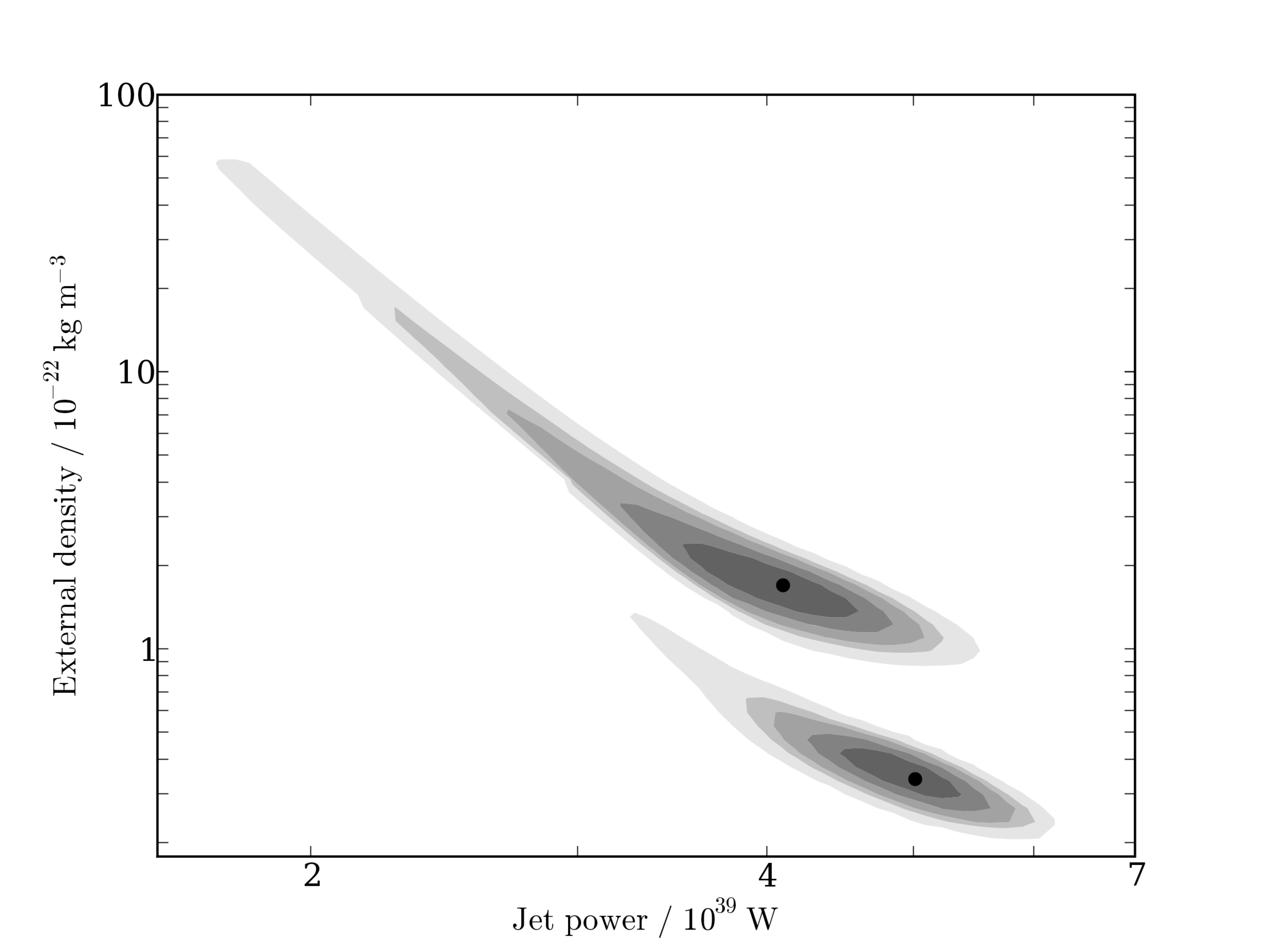}
\caption{Confidence contours for the free model parameters jet power, $Q$, and environment density, $\rho$, for the model of the outer lobes of B\,1450+333 with $\beta=1.5$ and $k=100$. Upper contours for the north outer lobe, lower contours for the south outer lobe. The contours from the outside inwards are for confidence levels 99\%, 95\%, 90\%, 75\% and 50\%. Black dots show the location of the best fit parameters summarised in Table \ref{1450mod}.}
\label{1450powerdensity}
\end{figure}

The uncertainties in the flux measurements at the frequencies used to fit the model spectra give rise to uncertainties in the free model parameters. The density in the host galaxy environment is associated with the largest uncertainty. Fig.~\ref{1450powerdensity} shows the confidence contours in the jet power -- density model plane. The contours for the outer lobes overlap considerably with respect to the range in jet power preferred by the model, but the model indicates a higher density to the north of the host galaxy. Note that in this and the following figures the confidence contours would cover smaller ranges for the smaller errors on the flux measurements suggested above. The quality of the data makes it particularly difficult to constrain the source parameters; however, for further discussion of the discrepancies between model parameters for each pair of lobes, see \citet{mjs09}.

\begin{figure}
\includegraphics[width=8.45cm]{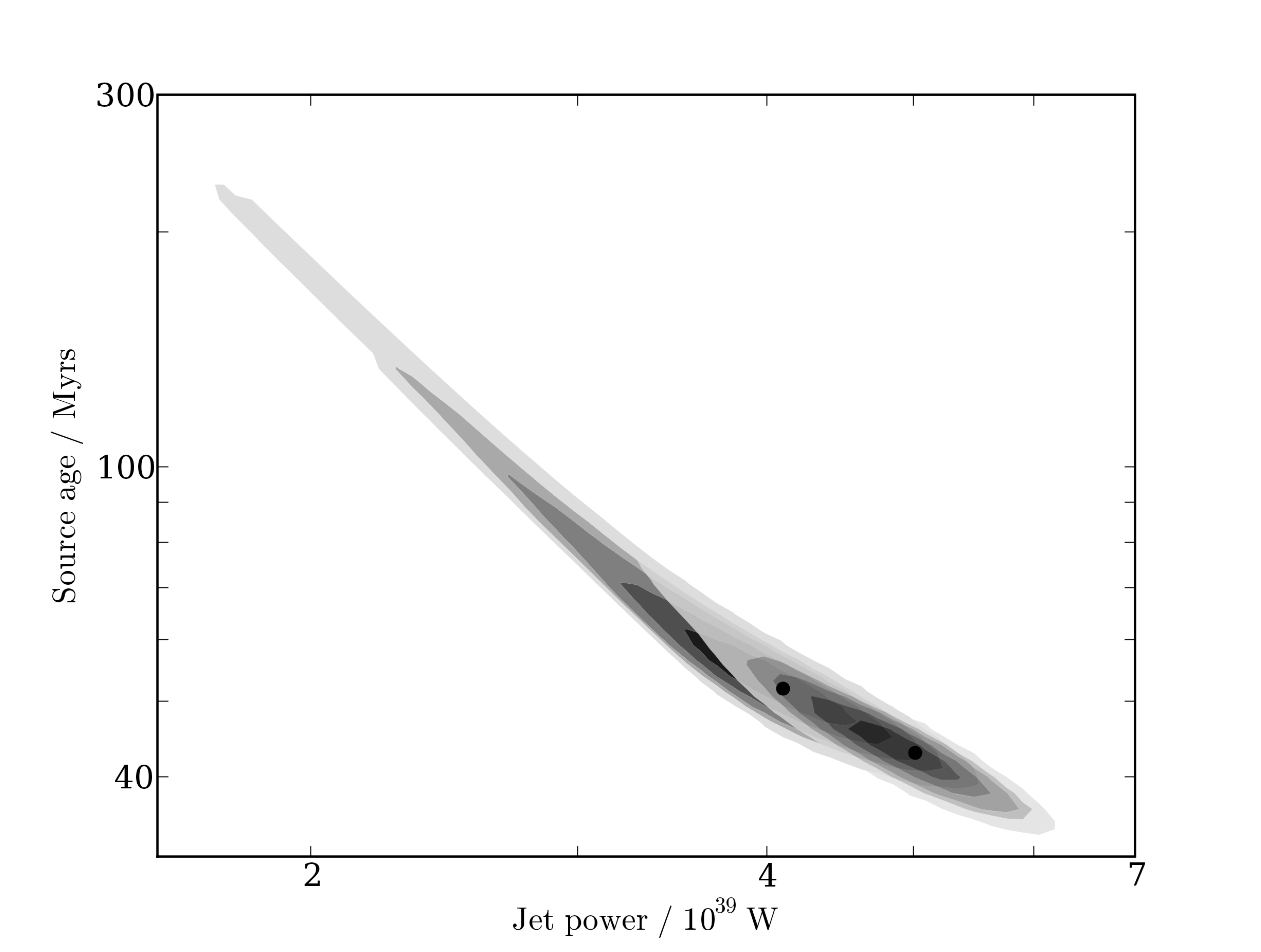}
\caption{Same as Fig.~\ref{1450powerdensity}, but for jet power, $Q$, and source age, $t$. The contours more extended towards lower jet powers show the model parameters for the north outer lobe. The other set of contours shows the south outer lobe.}
\label{1450powertime}
\end{figure}

Fig.~\ref{1450powertime} shows the confidence contours in the jet power -- source age plane. Both $Q$ and $t$ should be the same for both lobes and it is reassuring that the overlap of the confidence contours in this plane is large. The model constrains the source age and the jet power roughly to within a factor three. 

\begin{figure}
\includegraphics[width=8.45cm]{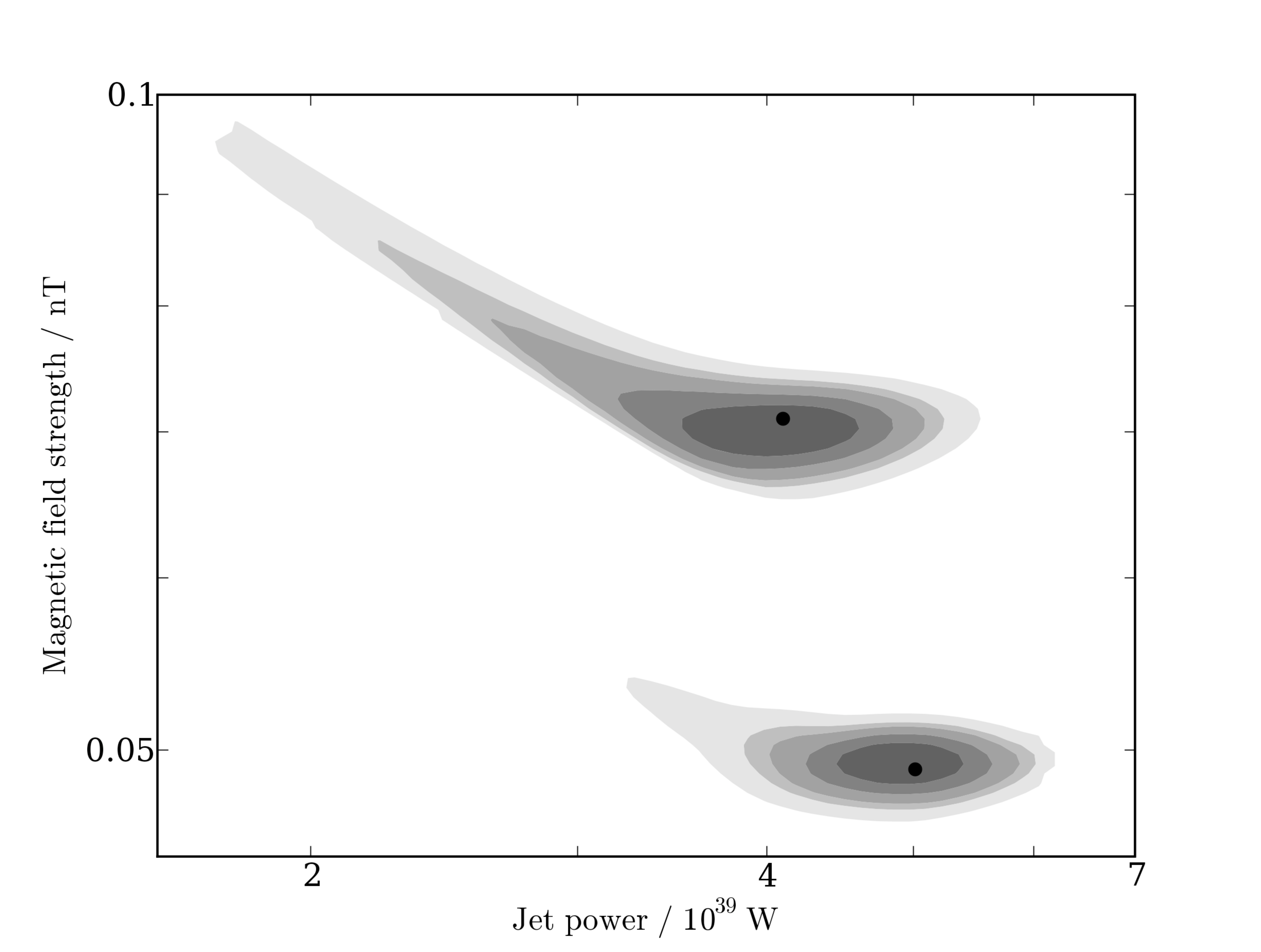}
\caption{Same as Fig.~\ref{1450powerdensity}, but for jet power, $Q$, and magnetic field strength in the lobe, $B_{\rm l}$. The upper contours show the parameters for the north outer lobe, the lower contours for the south outer lobe.}
\label{1450powerfield}
\end{figure}

Finally, the confidence contours in the jet power -- magnetic field strength plane are shown in Fig.~\ref{1450powerfield}. As mentioned above, the model strongly constrains the strength of the magnetic field in the lobes. This demonstrates that the strength of the magnetic field and hence the pressures in the lobes are the best constrained parameters of the model. The values of $B_{\rm l}$ and $p_{\rm l}$ do not change much even for the assumption of $\beta =0$ or for changes of the other set model parameters. 

\subsubsection{Inner lobes}

The observed radio spectra of the inner lobes of B\,1450+333 are well described by the power-laws 
\begin{equation}
F_{\nu} = 6 \left( \frac{\nu}{\rm MHz} \right)^{-0.71} \, {\rm Jy},
\label{1450fnunorth}
\end{equation}
for the northern inner lobe and 
\begin{equation}
F_{\nu} = 5.8 \left( \frac{\nu}{\rm MHz} \right)^{-0.84}\,{\rm Jy}
\label{1450fnusouth}
\end{equation}
for the southern inner lobe. Fig.~\ref{1450spec} shows these fits translated to the source restframe. We can now apply the modified standard FRII model described in section \ref{stand} to the observations by setting $m=2.42$ for the northern inner lobe and $m=2.68$ for the southern inner lobe. This assumes that the radiative energy losses of the relativistic electrons in the inner lobes are not significant enough to affect the spectra. 

As discussed in section \ref{stand}, the set of model parameters $Q$, $\rho_{\rm i}$ and $t_{\rm i}$ cannot be uniquely fitted to the observations in this case. Hence we take the value of $Q$ from each of the models fitted to the data of the respective outer lobe and calculate $\rho_{\rm i}$ and $t_{\rm i}$ for this value of $Q$ from equations (\ref{innert}) and (\ref{innerQ}). Here, we make the implicit assumption that the power of the jets that inflate the inner lobes is equal to the power of the jets that produced the outer lobes, although it is indeed possible that this assumption is not valid \citep[see e.g.][]{sbrlk00, mjk10}. We also assume that the density distribution inside the outer lobes, which provide the environment the inner lobes are expanding into, is uniform, i.e. $\beta = 0$. The results are summarised in Table \ref{1450mod}.

Clearly, the standard FRII model requires young ages for the inner lobes, $t_{\rm i}$. All values of $t_{\rm i}$ are consistent with the requirement that the inner lobes must have been inflated since the jet flows to the outer lobes stopped, i.e. in every case $t_{\rm i} \le t - t_{\rm j}$. The young ages also require the advance speed of the leading edge of the inner lobes to be fast. For the models of the outer lobes with $k=0$, these advance speeds averaged over the age of the inner lobes lie between $0.1\,c$ and $0.7\,c$. However, for the preferred models with $k=100$ the average advance speeds become superluminal. Such an unphysical situation can only be avoided if the power of the jets inflating the inner lobes drops compared to the jet power associated with the outer lobes. 

Another problem for the standard FRII model arises from the densities required by the model for the inner lobes. Assuming a jet power $Q$, the model makes a prediction for the density of the gas surrounding the inner lobes, $\rho_{\rm i}$. Since the inner lobes expand inside the outer lobes, this density $\rho_{\rm i}$ must be equal to the density inside the outer lobes. Combining equations (\ref{outerdensity}) and (\ref{masstransport}) we can derive an estimate for the density inside the outer lobes, $\rho_{\rm l}$. The density $\rho_{\rm l}$ arises solely from the material that was transported along the jets that inflated the outer lobes. It does not take into account a possible `contamination' of the outer lobes by mixing in material across the outer lobe boundaries. For simplicity Table \ref{1450mod} lists the values for $\rho_{\rm l} \left( \xi = 1 \right)$ after setting the factor
\begin{equation}
\xi = \frac{{\cal R}\,_{\rm j}}{\gamma_{\rm j}^2 +{\cal R}\,_{\rm j} \left( \gamma_{\rm j}^2 - 1 \right)}
\end{equation}
equal to unity. For a self-consistent model without mixing across the outer lobe boundary we require $\rho_{\rm i} = \rho_{\rm l}$. It is straightforward to show that this results in 
\begin{equation}
{\cal R}\,_{\rm j} = \frac{\gamma_{\rm j}^2}{1 - \gamma_{\rm j}^2 +\rho_{\rm l} \left( \xi =1 \right) / \rho_{\rm i}}.
\end{equation}
By definition ${\cal R}\,_{\rm j} \ge 0$ and so we find a limit on the bulk Lorentz factor of the jet
\begin{equation}
\gamma_{\rm j} \le \sqrt{\frac{\rho_{\rm l} \left( \xi = 1\right)}{\rho_{\rm i}} + 1 }.
\label{gamj}
\end{equation}

The models with $k=0$ predict $\rho_{\rm l} \left( \xi = 1 \right) \ll \rho_{\rm i}$. While for models with $k=100$ the two density estimates are more comparable, we argued above that in these cases we expect a reduced power for the jets inflating the inner lobes which implies a higher value for $\rho _{\rm i}$ more comparable with that for the models with $k=0$. Therefore all models predict rather slow jets with $\gamma_{\rm j} \sim 1$, dominated by the transport of kinetic energy. Such slow jet velocities are unlikely and so to make the standard FRII model work we need to invoke significant mixing of high density gas from the source environment into the outer lobes. This is of course consistent with the findings of \citet{ksr00}, who applied the same model to DDRGs.

We now turn to the bow shock model for the inner lobes. To calculate the pressure inside the inner lobes, $p_{\rm s}$, as predicted by the dynamical model described in Section \ref{bowmod}, we use the model for the outer lobes with $\beta=1.5$ and $k=100$. This sets the jet power, $Q$, the volume of the outer lobes, $V_{\rm l}$, the duration of the jet flow inflating the outer lobes, $t_{\rm j}$, and the pressure inside the outer lobes, $p_{\rm l}$. From the observed geometry of the inner lobes we determine the area of the leading surface of the bow shock, $A_{\rm b}$. We assume a simple cylindrical geometry so that $A_{\rm b} = \pi \left[ D / \left( 2 A \right) \right]^2$. For given values of ${\cal R}\/_{\rm j}$ we can now calculate predictions for the pressure inside the inner lobes, $p_{\rm s}$, as a function of the Lorentz factor of the jet, $\gamma_{\rm j}$.

\begin{figure}
\includegraphics[width=8.45cm]{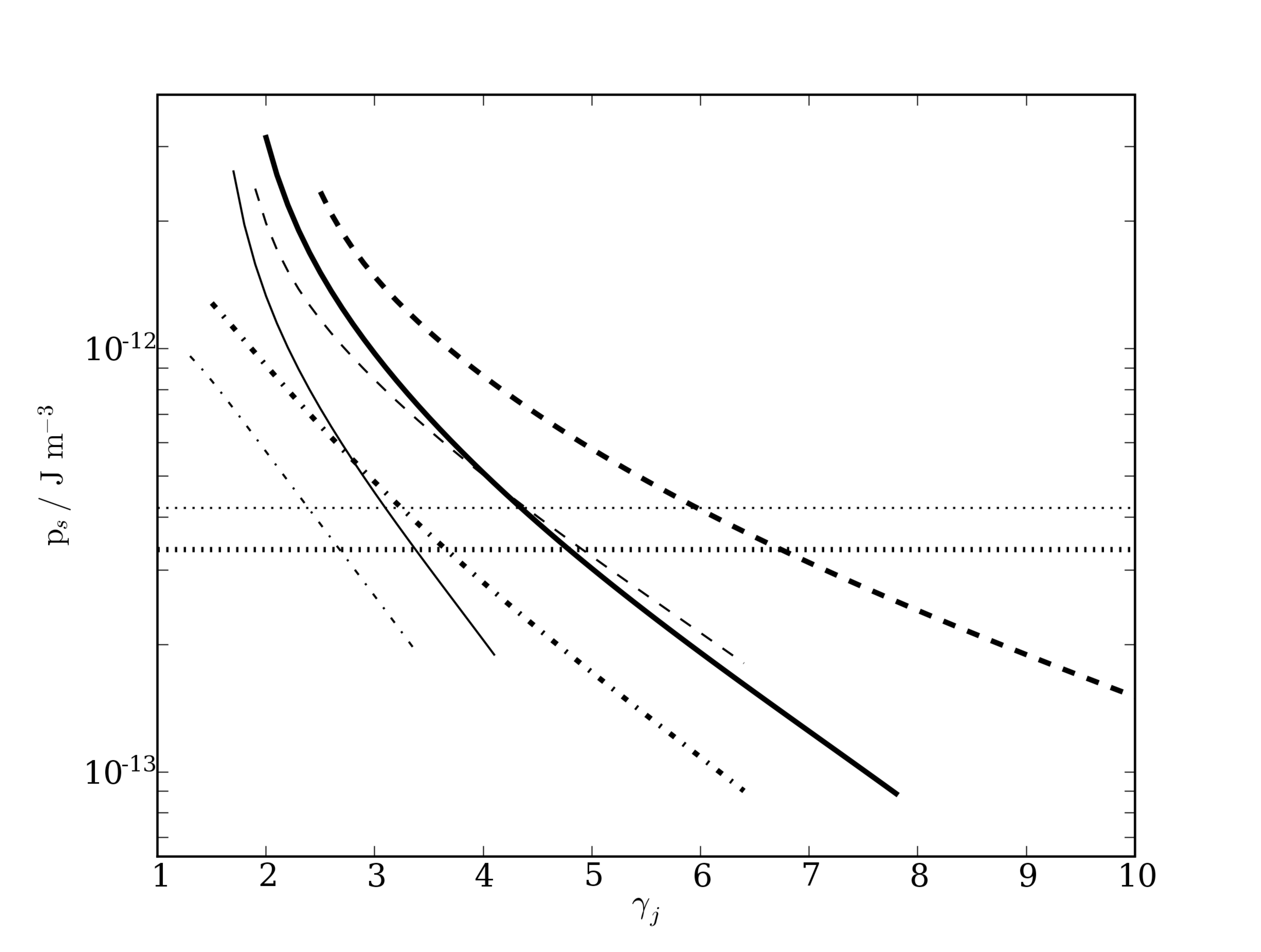}
\caption{The pressure in the inner lobes of B\,1450+333 predicted by the bow shock model as a function of the Lorentz factor of the jet. Light lines show the results for the north inner lobe, heavy lines for the south inner lobe. The dashed lines show results for relativistic jets dominated by the internal energy of the jet material, i.e. ${\cal R}\/_{\rm j} =0$, while dash--dotted lines show rest mass dominated jets with ${\cal R}\/_{\rm j} \rightarrow \infty$. Solid lines show the results for balanced jets, ${\cal R}\/_{\rm j} =1$. The horizontal dotted lines show the pressure inside the inner lobes derived from observed radio flux assuming a cylindrical geometry of the emission region and minimum energy conditions and setting $k=10$.}
\label{1450innerpressures}
\end{figure}

Fig.~\ref{1450innerpressures} shows the results for three values of ${\cal R}\/_{\rm j}$. Relativistic jets dominated by the internal energy of the jet material, i.e. ${\cal R}\/_{\rm j}=0$, predict the highest pressure inside the inner lobes for a given jet Lorentz factor, while the cold, matter dominated jets with ${\cal R}\/_{\rm j} \rightarrow \infty$ result in the lowest pressures. The endpoints of the curves at small $\gamma_{\rm j}$ are determined by the requirement that the shocked material behind the bow shock cannot move faster in the lobe restframe than the jet material itself. At large $\gamma_{\rm j}$ the curves terminate where equation (\ref{ps}) implies that there is no bow shock formed ahead of the jets, i.e. $p_{\rm s} = p_{\rm l}$. 

For a fixed jet power, $Q$, and value of ${\cal R}\/_{\rm j}$ the jet cross section, $A_{\rm j}$, is a decreasing function of $\gamma_{\rm j}$ (see equation \ref{area}). In other words, a faster jet achieves the same energy transport rate through a smaller cross section. Hence, equation (\ref{betab}) predicts a slower advance velocity for the bow shock, $\beta_{\rm b}$, for increasing $\gamma_{\rm j}$. This in turn reduces the compression ratio of the bow shock and the decreasing predicted pressure inside the inner lobes seen in Fig.~\ref{1450innerpressures}. Also, a slower bow shock implies a slower velocity for the shocked material behind the bow shock ahead of the jet. At the endpoints of the lines for large $\gamma_{\rm j}$, the shocked material is almost at rest inside the lobe restframe. Accordingly, the bulk velocity of the jet relative to the shocked lobe material increases from left to right in Fig.~\ref{1450innerpressures}. It is unlikely that the jet material can decelerate without forming a strong shock when this relative velocity is large. The absence of observed hotspots in the inner lobes and the problems of the standard FRII model discussed above therefore argue for solutions towards lower values of $\gamma_{\rm j}$.

Assuming minimum energy conditions inside the inner lobes, we can derive an independent estimate for the strength of the magnetic field \citep[e.g.][]{ml94}. We assume a cylindrical geometry of the inner lobes to calculate their volume. We have argued above that the radio spectrum is well fitted by the power-laws given in equations (\ref{1450fnunorth}) and (\ref{1450fnusouth}). The estimate of the minimum energy magnetic field, $B_{\rm min}$, should therefore not be significantly affected by energy losses of the radiating electrons and we use the spectral slope of the fitted power-laws. We derive two values for $B_{\rm min}$ for each of the inner lobes. The `conservative' value assumes that there is no radio emission outside the redshift-corrected frequency band across which we have observations of the inner lobes (see Table \ref{1450obs}). The `traditional' value assumes that the radio spectrum extends between the essentially arbitrary limits 10\,MHz to 100\,GHz. The traditional value will always exceed the conservative one. In practice the conservative and traditional values for $B_{\rm min}$ are not very different. For the inner north lobe we find $B_{\rm min}$ equal to 0.22\,nT or 0.32\,nT, respectively. For the inner south lobe we get 0.18\,nT and 0.28\,nT. 

Fig.~\ref{1450innerpressures} also shows the pressures inside the inner lobes derived from the traditional values of $B_{\rm min}$ under the assumption that the non-radiating particles in the inner lobes store ten times more energy than the relativistic electrons, i.e. $k=10$. This is lower than in the outer lobes with $k=100$ which the bow shocks are expanding into. However, the pressure derived assuming minimum energy conditions in the inner lobes exceeds any pressure achievable by the bow shock. Reducing the value of $k$ across the bow shock implies that the shock channels energy preferentially to the relativistic electrons and the magnetic field. Some re-acceleration of electrons at the bow shock seems plausible, particularly since the observed spectral slopes of the inner lobes are different from those of the outer lobes, and so a small reduction of $k$ may occur. Note that $k$ must decrease more for higher $\gamma_{\rm j}$. Unless $k$ is reduced significantly, this again argues for moderate values of $\gamma_{\rm j}$. 

The bow shock model can explain the observations of the inner lobes without the need to invoke significant mixing of the outer lobes with the surrounding heavier gas. To avoid the formation of a strong termination shock at the end of the jets, the model predicts moderate bulk jet velocities with $\gamma _{\rm j} \sim 2 \rightarrow 3$. The bow shocks driven by these jets then propagate through the outer lobes with velocities comparable to the bulk jet velocities. Despite their considerable size, the inner lobes are young in this model with ages of only a few $10^5$\,years consistent with the absence of any spectral ageing signature from the observed spectra. These young ages are also in agreement with the time available for the formation of the inner lobes, $t-t_{\rm j}$, as predicted by the model of the outer lobes. 

\subsection{B\,1834+620}

Table \ref{1834obs} summarises the observational data used to constrain the models for the inner and outer lobes of B\,1834+620. As for B\,1450+333, we assume a typical error of 7\% on all flux measurements.

\begin{table}
\begin{threeparttable}
\begin{tabular}{lcccc}
\hline
\hline
{\bf B1834+620:}& \multicolumn{2}{c}{North} & \multicolumn{2}{c}{South}\\
& Outer & Inner & Outer & Inner\\
Lobe length, $D$ / kpc & 713\tnote{a} & 202\tnote{b} & 736\tnote{a} & 194\tnote{b}\\
Aspect ratio, $A$ & 12\tnote{a} & 14\tnote{b} & 13\tnote{a} & 16\tnote{b}\\
\hline
& \multicolumn{4}{c}{Flux densities / mJy}\\
\hline
0.612\,GHz & 573 & 309 & 357 & 422\\
0.845\,GHz & 435 & 228 & 259 & 298\\
1.40\,GHz & 275 & 144 & 177 & 198\\
4.85\,GHz & -- & 30.9 & -- & 51\\
8.46\,GHz & 53 & 19.4 & 32.3 & 29.1\\
\hline
\end{tabular}
\caption{Observed properties of the four lobes in the DDRG B\,1834+620 used in the modelling. All flux densities are taken from the compilation by \citet{sbrl00}. We do not use the flux density measurements at 1.395\,GHz and 1.435\,GHz quoted in that paper as these observations are suspected to resolve out some of the flux.}
\label{1834obs}
\begin{tablenotes}
\item [a] VLA in B-array at 8.5\,GHz
\item [b] VLA in A-array at 4.8\,GHz
\end{tablenotes}
\end{threeparttable}
\end{table}

\subsubsection{Outer lobes}

\begin{figure}
\includegraphics[width=8.45cm]{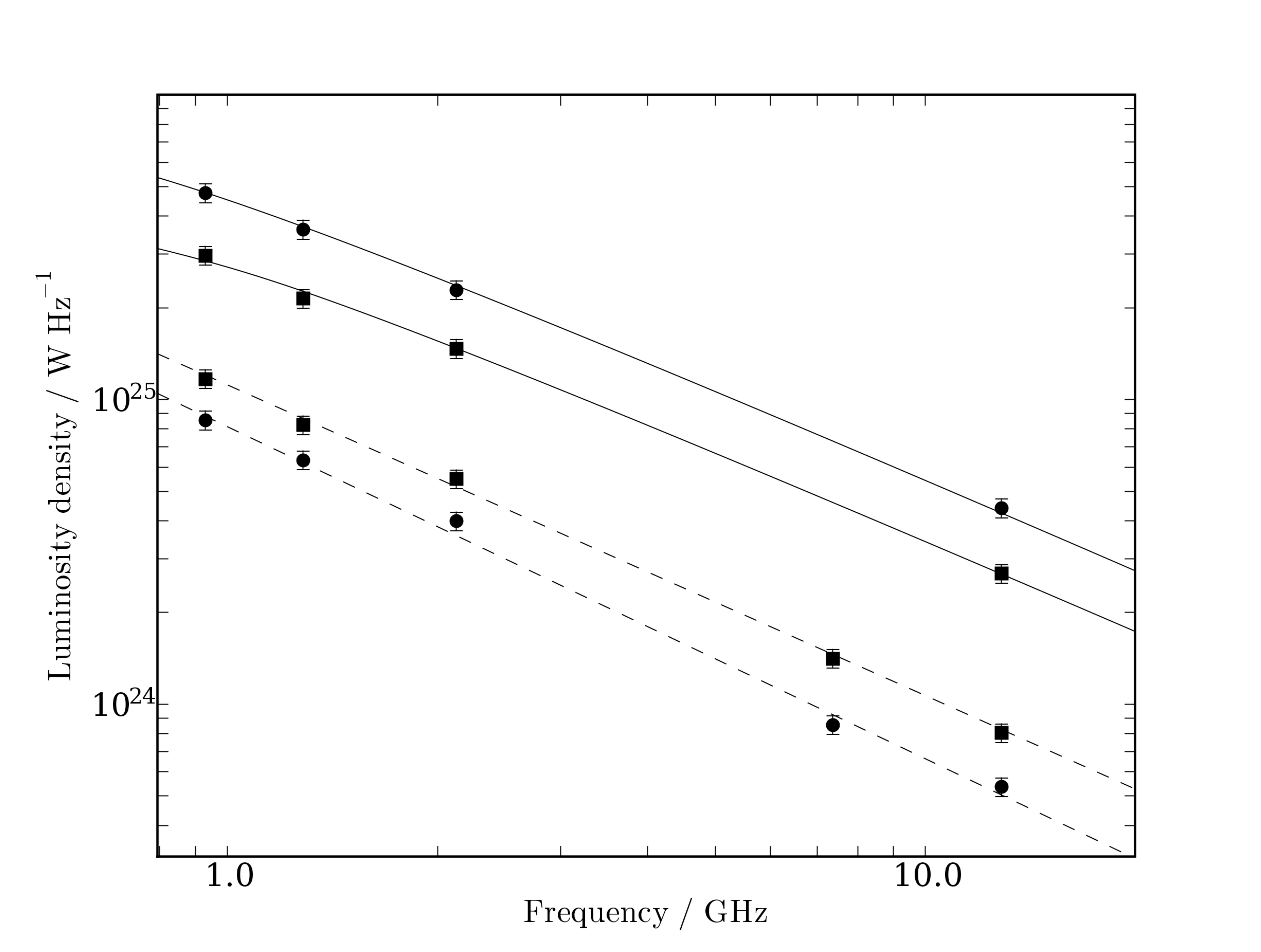}
\caption{Comparison of model spectra with observations for B\,1834+620 in the source restframe. Circles show the data for the northern lobes, squares are for the southern lobes. The upper data points and lines are associated with the outer lobes while the lower data points and lines show the inner lobes. The solid lines show the model predictions for the outer lobes, the dashed lines show the power-law fits to the spectra of the inner lobes. The luminosities of the inner lobes have been divided by a factor 3 for clarity.}
\label{1834spec}
\end{figure}

For the outer lobes we have flux measurements at four observing frequencies. It follows from Section \ref{KDA} that there is only one degree of freedom in the model fit. Again we attempt fits with three models differing in their assumptions about the slope of the external density profile, $\beta$, and the ratio of the energies stored in non-radiating particles and relativistic electrons, $k$. We tried models with $\beta =1.5$ and $k=0$, $\beta =0$ and $k=0$ and $\beta =1.5$ and $k=10$. The fitting results are summarised in Table \ref{1834mod} and the model spectrum for $\beta =1.5$ and $k=0$ is compared to the observations in Fig.~\ref{1834spec}.

\begin{table*}
\begin{tabular}{lcccccc}
\hline
\hline
 & \multicolumn{2}{c}{$k=0$, $\beta=1.5$} & \multicolumn{2}{c}{$k=0$, $\beta =0$} & \multicolumn{2}{c}{$k=10$, $\beta=1.5$}\\
 & North & South &North & South & North & South\\
 \hline
 $Q / 10^{37}\,{\rm W}$ & 89 & 78 & 72 & 63 & 620 & 540\\
 $t / {\rm Myr}$ & 28 & 25 & 28 & 25 & 26 & 24\\
 $t_j / {\rm Myr}$ & 26 & 23 & 26 & 23 & 24 & 21\\
 $\rho / 10^{-24}\,{\rm kg}\,{\rm m}^{-3}$ & 210 & 130 & 0.075 & 0.045 & 1200 & 770\\
 $\chi ^2$ & 0.62 & 0.95 & 2.3 & 3.0 & 0.49 & 0.90\\[1ex]
 $p_{\rm l} / 10^{-15}\,{\rm J}\,{\rm m}^{-3}$  & 18 & 13 & 21 & 15 & 120 & 84\\
 $B_{\rm l} / {\rm nT}$ & 0.17 & 0.14 & 0.18 & 0.16 & 0.17 & 0.14\\
 $\rho_{\rm l} \left( \xi = 1 \right) / 10^{-24}\,{\rm kg}\,{\rm m}^{-3}$ & $1.4 \times 10^{-4}$ & $1.2 \times 10^{-4}$ & $1.1 \times 10^{-4}$ & $9.3\times 10^{-5}$ & $9.0 \times 10^{-4}$ & $7.3 \times 10^{-4}$\\
 \hline
 $\rho_{\rm i} / 10^{-24}\,{\rm kg}\,{\rm m}^{-3}$ & 0.066 & 0.15 & 0.10 & 0.22 & 0.012 & 0.026\\
 $t_{\rm i} / {\rm Myr}$ & 2.8 & 3.3 & 3.5 & 4.1 & 0.84 & 0.97\\
\hline 
\end{tabular}
 \caption{Model parameters for the outer lobes of B\,1834+620 for the model spectra with the best fit. The model parameters for the inner lobes for the standard FRII model, $t_{\rm i}$ and $\rho_{\rm i}$, evolving inside the respective outer lobes are also given. For this source the values of $t_{\rm i}$ and $\rho_{\rm i}$ are calculated for a truncated radio spectrum. See text for a detailed discussion.}
 \label{1834mod}
 \end{table*}

Both models with $\beta =1.5$ achieve good agreement with the observational data. The model with $\beta =0$ provides a significantly weaker fit. The $\chi^2$ values for the models with $\beta =1.5$ are again below unity and may indicate that the error of 7\% assumed for the flux measurements may be too large. 

All fits use a power-law exponent $m=2$ for the energy distribution of the electrons when they are injected into the lobes. The observed spectral slope is steeper than $\alpha =0.5$ and so significant ageing affects the spectrum. However, the observed spectrum shows very little curvature and the flux measurements for the outer lobes can be adequately fitted with simple power-laws. In fact, the spectra can be fitted accurately by the standard FRII model for parameters avoiding much spectral ageing, if we assume that $m>2$. For example, we obtain an excellent fit, $\chi^2 = 0.38$, for the spectrum of the north outer lobe by setting $m=2.5$. The problem with this solution is the very young age predicted by the model of $t=6$\,Myr. The outer lobes would have to expand with an average velocity of almost 0.4\,c and this requires an extremely high jet power in excess of $10^{40}$\,W. It is therefore unlikely that such a solution accurately describes the outer lobes of B\,1834+620. 

The model with $\beta =1.5$ and $k=0$ requires a density distribution in the source environment consistent with the range of properties in the \citet{jph07} galaxy group sample. In contrast to the situation for B\,1450+333, there is no need to increase $k$ in order to allow for denser, more realistic source environments. The source density required for the model with $k=10$ is about two times higher than the upper end of the range of the sample of \citet{jph07}. Much higher values for $k$, comparable to the $k=100$ for B\,1450+333, would again result in unacceptably high jet powers and densities for the source environment. These results strongly suggest that $k$ is not very large in the outer lobes of B\,1834+620.

\begin{figure}
\includegraphics[width=8.45cm]{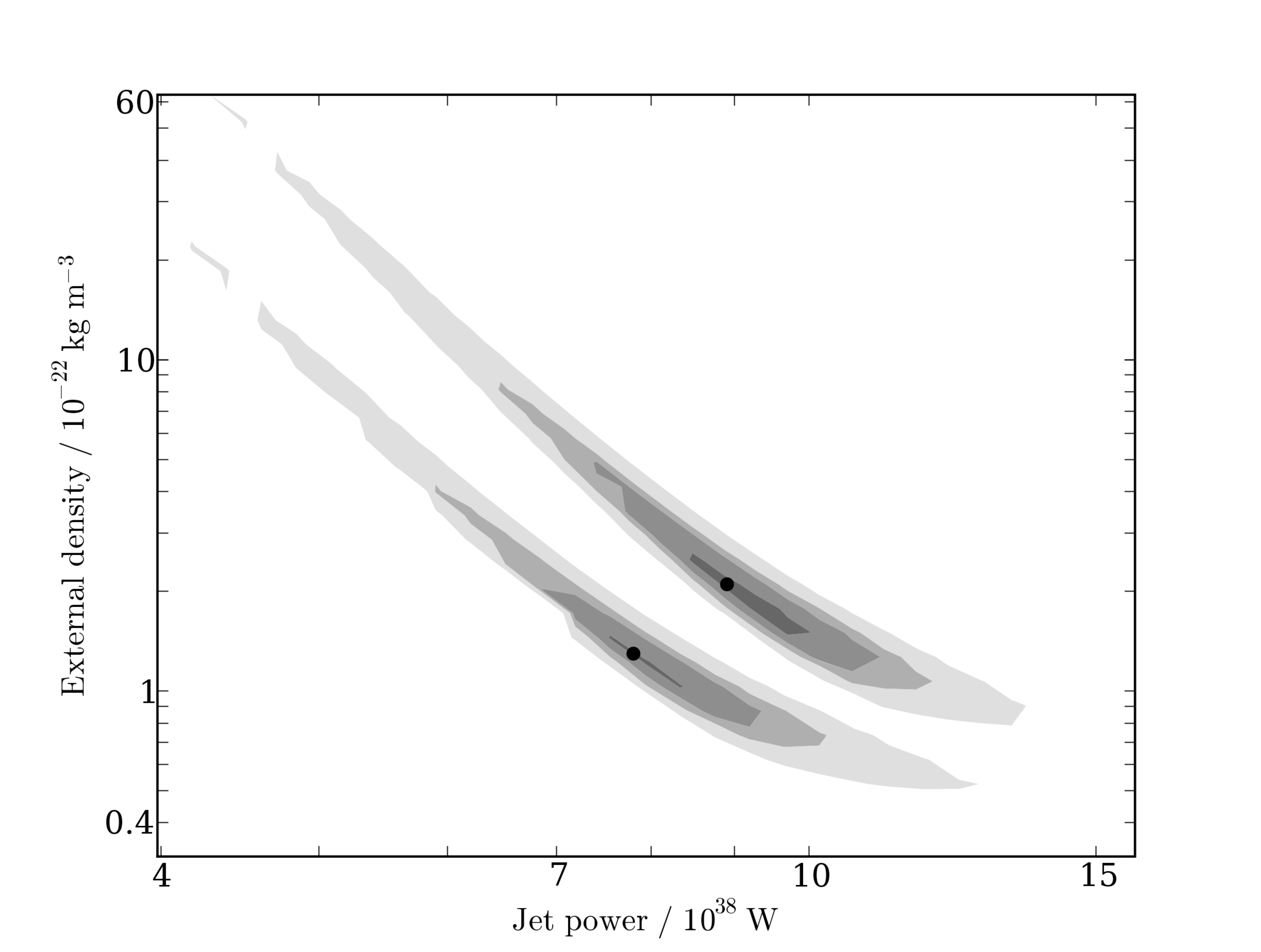}
\caption{Confidence contours for the free model parameters jet power, $Q$, and environment density, $\rho$, for the model of the outer lobes of B\,1834+620 with $\beta=1.5$ and $k=0$. Upper contours for the north outer lobe, lower contours for the south outer lobe. The contours from the outside inwards are for confidence levels 99\%, 95\%, 90\% and 75\%. Black dots show the location of the best fit parameters summarised in Table \ref{1834mod}.}
\label{1834powerdensity}
\end{figure}

\begin{figure}
\includegraphics[width=8.45cm]{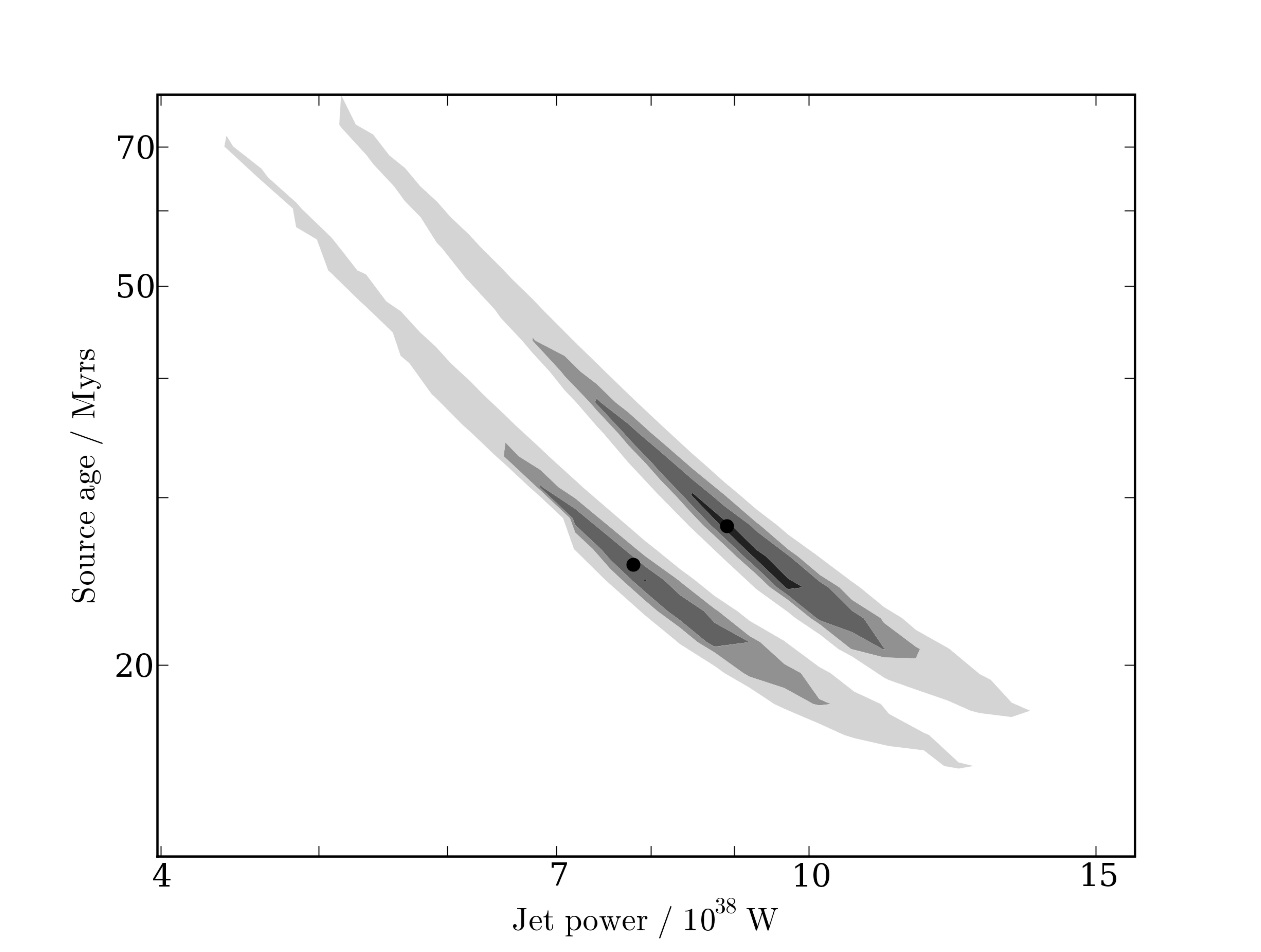}
\caption{Same as Fig.~\ref{1834powerdensity}, but for jet power, $Q$, and source age, $t$. The upper contours show the model parameters for the north outer lobe. The lower set of contours shows the south outer lobe.}
\label{1834powertime}
\end{figure}

\begin{figure}
\includegraphics[width=8.45cm]{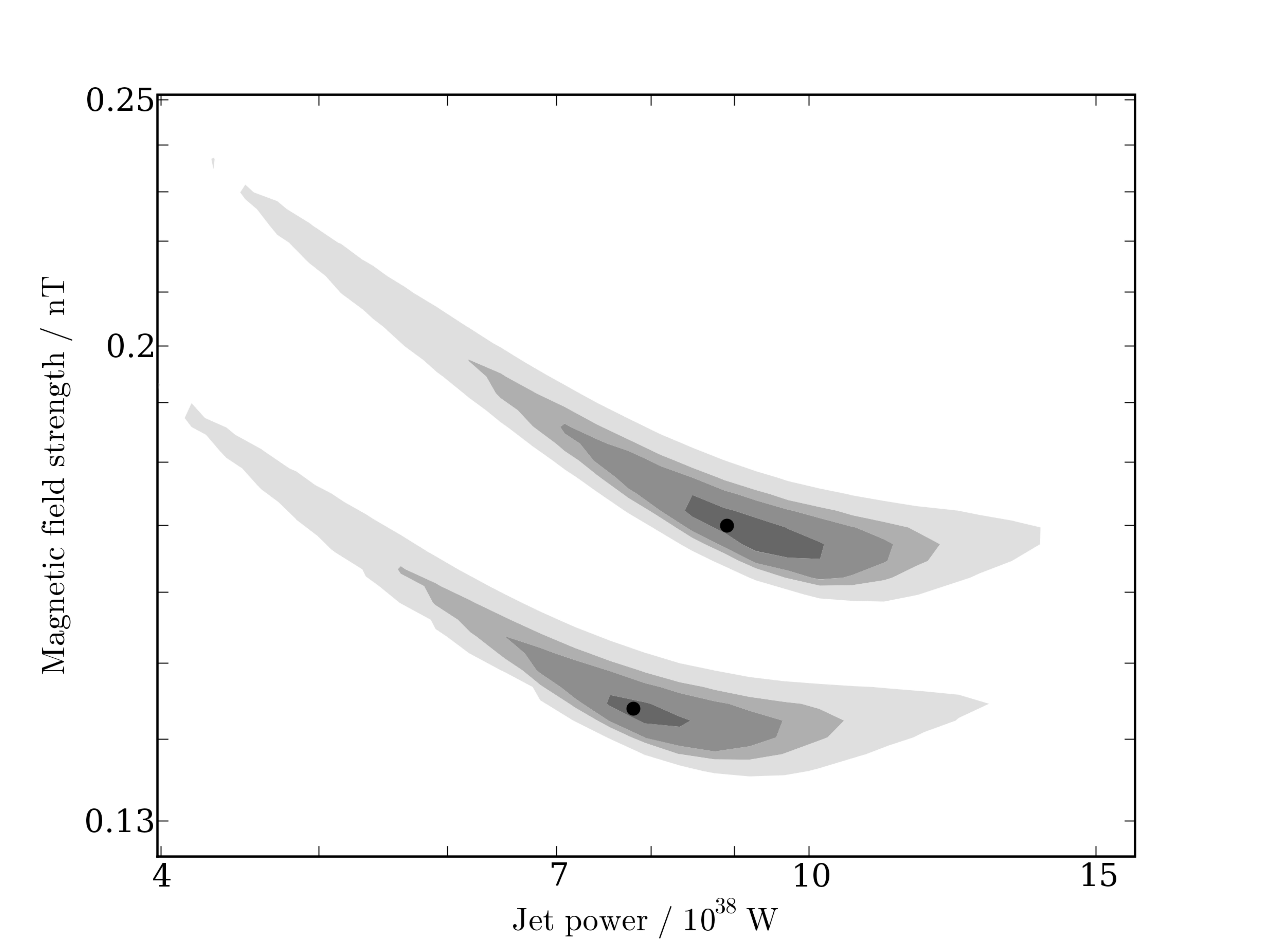}
\caption{Same as Fig.~\ref{1834powerdensity}, but for jet power, $Q$, and magnetic field strength in the lobe, $B_{\rm l}$. The upper contours show the parameters for the north outer lobe, the lower contours for the south outer lobe.}
\label{1834powerfield}
\end{figure}

Figs.~\ref{1834powerdensity}, \ref{1834powertime} and \ref{1834powerfield} show the confidence contours on the model parameters for the assumed error on the flux measurements. Again, the smaller errors argued for above would decrease the range of the confidence contours further. These plots are analogous to Figs.~\ref{1450powerdensity}, \ref{1450powertime} and \ref{1450powerfield} for B\,1450+333. The model parameters are constrained to a similar degree for B\,1834+620. However, in Fig.~\ref{1834powertime} the contours for the two outer lobes do not overlap. We would expect that the plotted parameters, the jet power and the source age, should be the same for both lobes on either side of the source. While this discrepancy is certainly visible in the figure, it is not very large. It is impossible to trace the outer lobes of B\,1834+620 all the way back to the core in our radio maps. Hence our possible overestimate of the aspect ratio of one of the lobes discussed in section \ref{KDA} may cause this problem.

An important constraint arising from the model of the outer lobes is the very short time available for the inner lobes to expand. They can only be inflated after the jet flow to the outer lobes ceases at $t_{\rm j}$. Hence the age of the inner lobes must be less than $t-t_{\rm j} \sim 2$\,Myr. Given the large size of the inner lobes of B\,1834+620, this implies a very fast expansion. We will see in the next section that this tight constraint makes the case for the bow shock model for the inner lobes even stronger than in the case of B\,1450+333. 

\subsubsection{Inner lobes}

The observed spectra of the inner lobes of B\,1834+620 are steeper than any of the other spectra discussed in this paper. The spectrum of the north inner lobe is well fitted by
\begin{equation}
F_{\nu} = 350 \left( \frac{\nu}{\rm MHz} \right)^{-1.09}\,{\rm Jy},
\end{equation}
while the spectrum of the south inner lobe is close to
\begin{equation}
F_{\nu} = 300 \left( \frac{\nu}{\rm MHz} \right)^{-1.02}\,{\rm Jy}.
\end{equation}
Because of its flatter spectral slope, the southern lobe is brighter in the observed range. This can be seen in Fig.~\ref{1834spec} where the fits are compared to the data in the source restframe. Note that the luminosities of the inner lobes are scaled by a factor 3 in this figure for clarity.

We apply the standard FRII model to the inner lobes of B\,1834+620 in the same way as in the case of B\,1450+333 with one important difference. From the power-laws fitted to the spectra we find that we need to set $m=3.18$ for the north inner lobe and $m=3.04$ for the south inner lobe. Hence the bulk of the energy in electrons in the model is stored in those electrons with the lowest Lorentz factors around $\gamma_{\rm min}$. So far we have used $\gamma _{\rm min} = 1$ and so those electrons storing the bulk of the energy do not contribute to the radio emission. As a consequence the standard FRII model predicts ages in excess of $10^8$\,years for the inner lobes, because the jets need to deliver a vast amount of energy to the inner lobes. The problem is solved by increasing the low energy cut-off of the electron energy distribution such that $\gamma _{\rm min} =1000$. The introduction of such a high cut-off leads to the truncation of the model radio spectrum. In fact, we cannot set a value for $\gamma _{\rm min}$ much larger than the 1000 used here or this cut-off will start affecting the model spectrum in the observed range. We did not introduce an equivalent high cut-off in the models for B\,1450+333 as for the flatter spectra of this source such a change makes little difference to the modelling results. Also, a further decrease in the age of the inner lobes of B\,1450+333 exacerbates the problems noted above. 

The results of the modelling are given in Table \ref{1834mod}. For this object the ages predicted for the inner lobes are too old for $k=0$ to be consistent with the properties of the outer lobes, despite the introduction of a large value for $\gamma _{\rm min}$. In all cases $t_{\rm i} > t - t_{\rm j}$ and so for $k=0$ we cannot build a self-consistent model combining standard FRII models for both the outer and the inner lobes. For $k=10$ the ages predicted for the inner lobes are smaller than $t-t_{\rm j}$. However, in this case the average expansion speed of the inner lobes approaches significant fractions of the speed of light. It is therefore unlikely that the jets inflating the inner lobes will terminate in strong shocks as required by the standard FRII model. 

As in the case of B\,1450+333 the densities required inside the outer lobes to explain the inner lobes are at least an order of magnitude higher than the model for the outer lobes predicts. Unless there is significant contamination of the outer lobe with dense gas from the source environment, equation (\ref{gamj}) then implies that the bulk speed of the jet must be very low. We conclude that the standard FRII model for the inner lobes of B\,1834+620 can only work, if extensive mixing of gas across the lobe surface has occurred. This is consistent with our findings for B\,1450+333 and \citet{ksr00}.

\begin{figure}
\includegraphics[width=8.45cm]{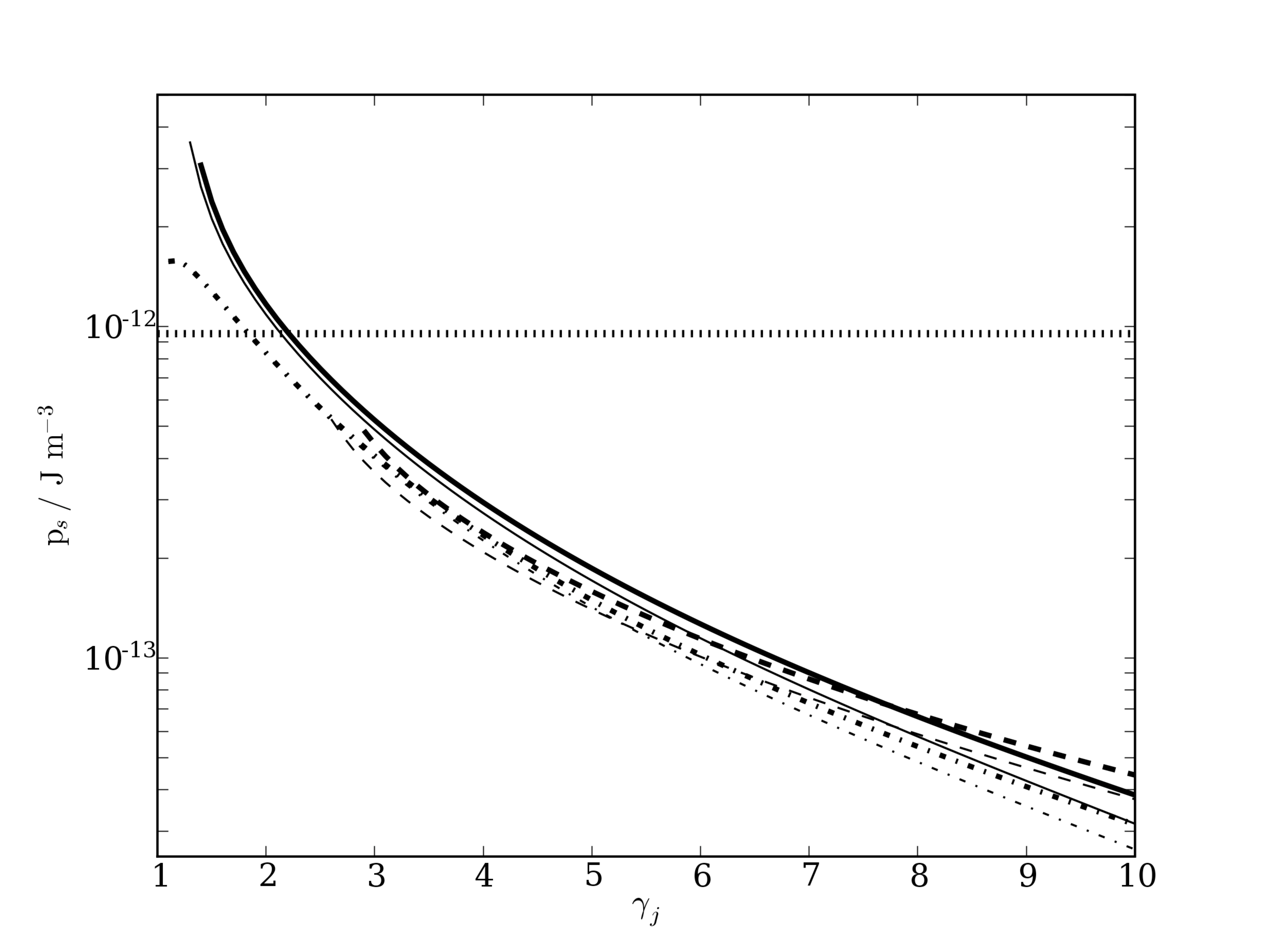}
\caption{Same as Fig.~\ref{1450innerpressures}, but for B\,1834+620. Note that the identical horizontal dotted lines here are calculated for $k=0$.}
\label{1834innerpressures}
\end{figure}

Next, we apply the bow shock model to the inner lobes of B\,1834+620 in the same way as in the case of B\,1450+333, but for the model of the outer lobes with $\beta =1.5$ and $k=0$. The pressure inside the inner lobes as predicted by the bow shock model are shown in Fig.~\ref{1834innerpressures}. There are several differences to the equivalent plot for B\,1450+333 shown in Fig.~\ref{1450innerpressures}. The curves for individual lobes, but for different values of ${\cal R}\/_{\rm j}$ are much closer to each other. The curves for the north inner lobe are also closer to the south inner lobe. This means that for a given jet speed the predicted pressures inside the lobes cover a small range independent of the form in which the jets transport the energy. For most values of $\gamma _{\rm j}$ the balanced jets with ${\cal R}\/_{\rm j} =1$ lead to the highest pressure. 

We also calculate the strength of the magnetic field corresponding to minimum energy conditions in the inner lobes. We find a conservative value of 0.53\,nT and a traditional value of 1.2\,nT for $B_{\rm min}$ in the north inner lobe. The corresponding values for the south inner lobe are 0.58\,nT and 1.2\,nT. The pressure in the inner lobes associated with the traditional values of $B_{\rm min}$ are shown as the dotted horizontal line in Fig.~\ref{1834innerpressures}. Here we have assumed that $k=0$ and so the horizontal lines represent a firm lower limit for $p_{\rm s}$. This is therefore a much stronger constraint than in the case of B\,1450+333. In this model we can rule out relativistic jets with ${\cal R}\/_{\rm j} =0$, because they do not allow high enough pressures inside the inner lobes. Also, the jet bulk Lorentz factor cannot be much greater than 2. The implied age of the inner lobes of roughly $0.8$\,Myr is consistent with the available time $t-t_{\rm j} \sim 2$\,Myr. Finally, the inner lobes are not consistent with large values of $k$. As with the outer lobes of B\,1834+620, we find that the inner lobes cannot contain an energetically very significant population of non-radiating particles. 

\section{Summary and outlook}

We find that the observed properties of the outer lobes of both, B\,1450+333 and B\,1834+620, can be modelled satisfactorily with the standard FRII model. The absence of hot spots from the radio maps is consistent with our interpretation that the jets in these sources have ceased to supply the outer lobes with energy. The steep radio spectra of the outer lobes are consistent with model solutions where the acceleration of relativistic electrons stopped a few Myr for B\,1834+333 to several Myr for B\,1450+333 ago. 

Unless the outer lobes of B\,1450+333 are embedded in an environment with an exceptionally low gas density, we find that they must contain a population of non-radiating particles that store a significant amount of the energy originally transported by the jets. For a ratio of energy stored in non-radiating particles to relativistic electrons, $k$, equal to 100 we find a model fit with a realistic gas density in the environment. In contrast for B\,1834+620 such a large value of $k$ would lead to an unrealistically high jet power and short source age. While setting $k=10$ for this source results in an acceptable model fit, the observed properties are also consistent with $k=0$, a negligible contribution to the total energy in the lobe by non-radiating particles. 

In both sources the time available for the inner lobes to be inflated, defined as the difference between the total source age, $t$, and the duration of the first jet activity creating the outer lobes, $t_{\rm j}$, is comparatively short. This, together with their considerable size, implies that the inner lobes must propagate rapidly through the outer lobes, independent of the specific model we apply to the inner lobes. 

In both sources the standard FRII model struggles to accommodate the observed properties of the inner lobes. The model is based on the formation of a strong shock at the end of the jets caused by ram pressure balance with the surrounding medium. The gas density inside the outer lobes is simply not high enough to effectively stop the jets unless their bulk velocity is very slow. The outer lobes may be `contaminated' by denser gas from their surroundings, but the processes leading to such a contamination are slow \citep{ksr00}. An alternative solution to make the standard FRII model consistent with the observations is to invoke a significant drop of the power of the jets inflating the inner lobes compared to the jets responsible for the outer lobes of the same source. While we cannot rule this out for B\,1450+333, it cannot explain the inner lobes of B\,1834+620 as their predicted age would increase and become even less consistent with the constraint provided by the outer lobes of this source. 

The bow shock model for the inner lobes interprets them as arising from the emission of relativistic electrons compressed and re-accelerated by the bow shock in front of the jets inside the outer lobes. In this model the jets in the inner lobes do not decelerate significantly, they do not form radio hot spots - the presence of hotspots in a DDRG would rule out the bowshock model, providing an observational test  - and the lobes are expected to expand fast. The model is consistent with the inner lobes in both sources. Although we cannot determine the bulk jet velocity from the models, our results suggest that the jet bulk Lorentz factors should be around 2. The constraints are much stronger for B\,1834+620 because of its larger inner lobes. In this source we can also rule out jets whose energy transport is dominated by the internal energy of a highly relativistic jet material. 

In the case of B\,1450+333 the bow shock model suggests a significant contribution of non-radiating particles to the total energy stored in the inner lobes. However, the value of $k$ is likely somewhat lower than for the outer lobes in this object. For B\,1834+620 the inner lobes as well as the outer lobes cannot contain non-radiating particles making a significant contribution to the total energy.

A combination of the standard FRII model for the outer lobes with the bow shock model for the inner lobes provides a self-consistent interpretation for both sources studied in this work. With this combined model we do not need to invoke a drop in the jet powers or a significant contamination of the outer lobes with dense gas as proposed in \citet{ksr00}. While we cannot rule out these processes, we can explain the observations without them. 

The bow shock model for the inner lobes predicts a very rapid propagation of the bow shock through the outer lobes. In both sources the working surface of the jets driving the bow shocks ahead of them should reach the tip of the outer lobes only about 2 to 3\,Myrs after they started their expansion at the source centres. At this point the source will develop new hotspots at the ends of the outer lobes and have the appearance of a `normal', although large, FRII-type radio galaxy. This may imply that many, if not all, large radio galaxies go through short-lived DDRG phases. We may even speculate about a general process acting in radio galaxies whereby the jet production is infrequently interrupted in the central AGN for short periods of time, say of order 1\,Myr.  The jet channels in the existing radio lobes collapse quickly ($\sim 10^4$ years, comparable with the sound-crossing time for the jet radius) and after the jets restart, they drive bow shocks through the old lobes. During this phase the source would be classified as a DDRG. After the bow shocks reach the tips of the old lobes, the source becomes a standard FRII-type radio galaxy once more. Due to the fast propagation of the bow shocks the DDRG phase is short-lived compared to the total source age. Also, it lasts longer in larger sources simply because it takes the bow shocks longer to propagate to the ends of already large lobes. This may explain why DDRGs are rare and are typically associated with sources with very large outer lobes. In this scenario, the triple source B\,0925+420 \citep{bks07} is an example of an object where the jet production is interrupted again before the bow shocks in front of the new jets have reached the ends of the very large outer lobes. If such jet interruption occurs commonly in radio galaxies, then it should be possible to identify DDRG-like structures also in smaller objects, such as the recently discovered 4C~02.27 (\citealt*{jsk09}). 


\section*{Acknowledgements}
The VLA is a facility of the National Radio Astronomy Observatory, which is operated by Associated Universities Inc., under cooperative agreement with the National Science Foundation. We thank the anonymous referee for detailed comments which significantly improved the paper.

\def\newblock{\hskip .11em plus .33em minus .07em}

\bibliography{crk}
\bibliographystyle{mn2e}

\end{document}